\documentclass[fleqn,usenatbib]{mnras}

\usepackage{newtxtext,newtxmath}
\usepackage{caption}
\usepackage{bigints}
\usepackage[english]{babel} % To obtain English text with the blindtext package
\usepackage{blindtext}
\usepackage{graphicx}	% Including figure files
\usepackage{longtable}
\usepackage{color,soul}
\usepackage[T1]{fontenc}
\usepackage{graphicx}	% Including figure files
\usepackage{amsmath}	% Advanced maths commands
\usepackage{orcidlink}
\DeclareRobustCommand{\VAN}[3]{#2}
\let\VANthebibliography\thebibliography
\def\thebibliography{\DeclareRobustCommand{\VAN}[3]{##3}\VANthebibliography}

\usepackage{graphicx}	% Including figure files
\usepackage{amsmath}	% Advanced maths commands

\title[SN 2023ixf Nebular Spectra]{The nebular spectra of SN 2023ixf: A lower mass, partially stripped progenitor may be the result of binary interaction}

\author[P. D. Michel]{Philip D. Michel$^{\orcidlink{0009-0005-9716-6502}{1}},$\thanks{E-mail: philip.michel@gmail.com}
{Paolo A. Mazzali}$^{\orcidlink{0000-0001-6876-8284}{1,2}},$
{Daniel A. Perley}$^{\orcidlink{0000-0001-8472-1996}{1}},$
{K-Ryan Hinds}$^{\orcidlink{0000-0002-0129-806X}{1}}$, and
{Jacob L. Wise}$^{\orcidlink{0000-0003-0733-2916}{1}}$ \\
$^{1}$ Astrophysics Research Institute, Liverpool John Moores University,
146 Brownlow Hill,
Liverpool,
L3 5RF,
UK\\
$^{2}$ Max-Planck-Institut f\"{u}r Astrophysik, Karl-Schwarzschild Straße 1, 85748 Garching, Germany
}

\pubyear{2024}

\begin{document}
\label{firstpage}
\pagerange{\pageref{firstpage}--\pageref{lastpage}}
\maketitle

% Abstract of the paper

\begin{abstract}
SN 2023ixf is one of the brightest Core Collapse Supernovae of the 21st century and offers a rare opportunity to investigate the late stage of a Supernova through nebular phase spectroscopy.  We present four nebular phase spectra from day +291 to +413 after explosion.  This is supplemented with high cadence early phase spectroscopic observations and photometry covering the first 500 days to investigate explosion parameters.  The narrow and blue-shifted nebular Oxygen emission lines are used to infer an ejected Oxygen mass of $<0.65$M$_\odot$, consistent with models of a relatively low mass ($M_{ZAMS}<15$M$_\odot$) progenitor.  An energy of 0.3 to $1.4  \times10^{51}$ erg and a light curve powered by an initial $^{56}$Ni mass of $0.049 \pm 0.005 $M$_\odot$ appear consistent with a relatively standard Type II explosion, while an incomplete $\gamma$-ray trapping (with timescale of $240\pm4$ days) suggests a lower ejecta mass.  Assuming a typical explosion, the broad Hydrogen and Calcium profiles suggest a common origin within a lower mass, partially stripped envelope.  Hydrogen emission broadens with time, indicating contribution from an additional power source at an extended distance; while the emergence of high velocity ($\sim$6,000 km s$^{-1}$) Hydrogen emission features (beginning around day +200) may be explained by Shock Interaction with a dense Hydrogen-rich region located at $\sim1.5 \times 10^{16}$cm.  Such envelope mass loss for a low mass progenitor may be explained through theoretical models of Binary interaction.
\end{abstract}
\begin{keywords}
Supernovae: SN 2023ixf -- Stars: Evolution
\end{keywords}
\section{Introduction}
\defcitealias{Silverman2017}{S17}
Core Collapse (CC) Supernovae (SNe) are the explosive end-lives of massive stars. Yet a direct link between the diverse range of observed SNe and their progenitors is an outstanding problem in SN research and stellar physics. During the nebular phase, the outer ejecta becomes optically thin, offering a brief window to observe the core left over from the explosion which can reveal information about the progenitor star.

Type II SNe (SNe-II) are distinguished from Type I by the presence of Hydrogen in their spectra \citep{Minkowski1941} and account for 75\% of all CC-SNe per unit volume \citep{10.1111/j.1365-2966.2011.18160.x, 2017PASP..129e4201S}.  The sub-type II-P refers to the Hydrogen rich sub-class who possess a “plateau” phase in their early light curve believed to originate from the recombination of Hydrogen which was ionised in the initial SN shock while the sub-type II-L show a more "linear" decline in magnitude brightness \citep{Barbon1979A&A}.   The transitional class IIb show early time Hydrogen features which disappear in later phases \citep{1988AJ.....96.1941F}. 

Several empirical studies \citep[eg][]{2012IAUS..279...34A, 2019MNRAS.488.4239P} illustrate that these subtypes fall within discreet lightcurve shapes suggesting distinct physical characteristics.  SNe II-L also typically reach higher maximum luminosity and show faster expansion velocities than II-P \citep{2014MNRAS.445..554F} indicating II-L SNe may have less massive Hydrogen envelopes to absorb the energy released during the early, optically thick phase of the explosion.  However, other studies have questioned a discrete classification system based purely on light curves \citep[eg.][]{2015ApJ...799..208S}.  Provided they are monitored for sufficient time, II-L SNe may also show a period of steep decline in brightness \citep{2014ApJ...786...67A} and therefore a plateau phase.  The discovery of intermediate examples showing properties of both II-L and II-P SNe \citep[eg.][]{2015MNRAS.448.2608V} clouds this distinction further.  So rather than distinct classes, CC-SNe may represent a continuum corresponding to the degree of Hydrogen and even Helium envelope mass loss of the progenitor \citep[eg][]{2006astro.ph..7422C}. Under such a model, II-P SNe are assumed to have retained their thick Hydrogen envelopes while Type I CC-SNe have been fully stripped of their Hydrogen.  Between the two extremes may lie a spectrum of Stripped-Envelope Supernovae (SE-SNe) classified by the extent of their envelope stripping.  

To understand the diversity of CC-SNe further, an understanding of their progenitor stars is required and serendipitous imaging of the pre-explosion site offers the most direct method of investigation.  The possible detection of binary systems at the site of stripped envelope IIb SNe \citep[eg][]{2002PASP..114.1322V, 2015MNRAS.454.2580M} may support the suggestion that binary interaction plays a part in envelope stripping  \citep{1992ApJ...391..246P, 2011A&A...528A.131C}.  Alternatively, strong stellar winds may lead traditional Red Supergiant (RSG) progenitors to evolve into Wolf-Rayet stars stripped of their Hydrogen envelopes \citep{1981A&A....99...97M}. Several progenitors of Type II-P SNe have been detected by this method, consistently suggesting that they originate from the core collapse of lower mass 8-18 M$_\odot$ RSGs \citep{2009ARA&A..47...63S} that have retained Hydrogen envelopes. One notable exception is the Blue Supergiant progenitor of the peculiar Type II-P SN 1987A which showed atypical features; most likely due to binary interaction of its progenitor \citep{1987ApJ...320..602K}.   Based on a more limited sample, II-L SNe appear to originate from slightly larger mass progenitors of $\sim 20$ M$_\odot$ \citep{2010ApJ...714L.254E, 2017suex.book.....B}.  However, stellar evolution models predict that starts upto 25 M$_\odot$ will evolve into RSGs \citep{2012A&A...537A.146E} and the absence of any observed progenitors above 20 M$_\odot$ has been referred to as the "RSG Problem" \citep{2015PASA...32...16S}.  Several arguments have been proposed to resolve the apparent absence of higher mass progenitors; including potential errors in the assumptions of progenitor luminosity and evolutionary paths \citep{2018MNRAS.474.2116D, 2025ApJ...979..117B}.  But this remains an active area of debate in the literature and an understanding of the possible link between progenitor masses and different SN explosions remains key to understanding the evolution of massive stars.

\begin{figure} 
\includegraphics[width=\columnwidth]{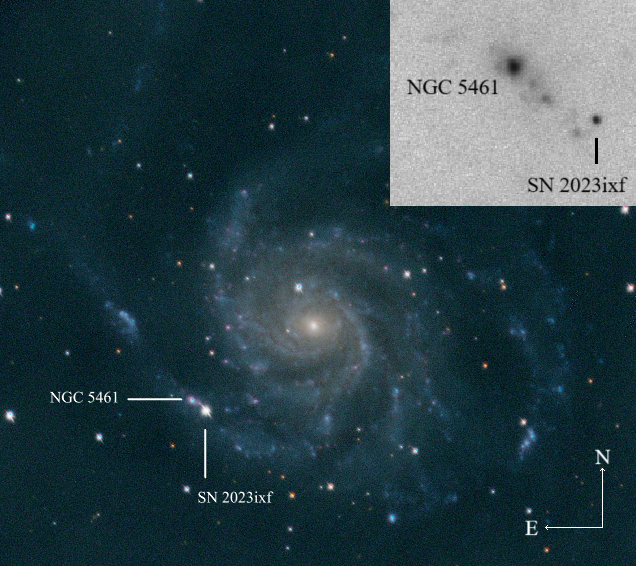}
    \caption{SN 2023ixf highlighted within the M101 spiral galaxy close to the H II region, NGC 5461.  Colour image taken with the author's telescope.  Monochrome inset shows a magnified unfiltered close-up of the region on day +291 taken with LT}
\label{m101image}
\end{figure}

After about 100 days, the SN enters the nebular phase where the outer layers become optically thin and the spectrum becomes dominated by forbidden emission lines \citep{1997ARAA..35..309F}. During this period, the spectrum can reveal important information about the core left over from the explosion and therefore offers an alternative method to investigate the progenitor star \citep{2011A&A...530A..45J}. 
 The luminosity of the light curve transitions to an exponential decay with time \citep{1945ApJ...102..309B} and spectral emission lines are primarily powered by the release of $\gamma$-rays and positrons from the radioactive decay of $^{56}$Co $\rightarrow$ $^{56}$Fe with an e-folding time of 111.3 days \citep{1969ApJ...157..623C}.  The primary power source for the luminosity of CC-SNe is therefore its parent isotope, $^{56}$Ni, which is created in the initial explosion and rapidly decays to $^{56}$Co in the early phase of the SN.   
 Assuming that all $\gamma$-rays are immediately absorbed and emitted through excitation and recombination, a linear decline in brightness of 0.98 mag (100 days)$^{-1}$ would be expected.  However, \citet{2014ApJ...786...67A} empirically found a wide range of SNe-II light curve decay rates with a mean drop in brightness of 1.47 mag (100 d)$^{-1}$ suggesting considerable diversity which may be influenced by the role of additional power sources, incomplete trapping of $\gamma$-rays or absorption from dust. 

While some Hydrogen rich Type II SNe appear to originate from compact RSGs, many show signs of circumstellar material (CSM) extending a considerable distance from the progenitor star \citep{2019A&A...631A...8H}.  As the SN ejecta expands, the shock wave's interaction with this CSM can become an increasingly important power source at later stages; providing Radio, Optical, X-Ray and UV emission \citep{2023A&A...675A..33D}.  This has even been observed in apparently Hydrogen poor SNe such as SN 2014C at delayed times of $> 100$ days suggesting interaction with Hydrogen rich material at extended distances of $>10^{16}$ cm \citep{2017ApJ...835..140M, 2019ApJ...887...75T} highlighting the remarkable diversity in the CSM structure of CC-SNe progenitors.   

Unfortunately, very few spectra exist in this fainter period due to the long exposure times required to obtain the necessary signal-to-noise ratio and only a small number of studies on individual SNe have included nebula phase spectra \citep[][hereafter, \citetalias{Silverman2017}]{Silverman2017}.  So nebular spectral analysis has historically been limited to a small sample of very nearby events.

SN 2023ixf therefore offers a fortunate opportunity to enrich the data in this space.  At a distance of $6.85 \pm 0.15$ Mpc \citep{2022ApJ...934L...7R}, it is the closest CC-SN to earth in recent decades. It reached a maximum apparent brightness of $m_V=10.8$ \citep{2023RNAAS...7..141S, 2023ApJ...954L..42J}, making it one of the brightest SNe observed from Earth in the 21st century and immediately became a high-profile target for amateur and professional astronomy using ground-based optical telescopes.

SN 2023ixf was discovered by \citet{2023TNSTR1158....1I} on 19 May 2023 at magnitude $m_V = 10.9$ in the spiral galaxy, M101 (at location $\alpha$ = 14:03:38.580, $\delta$ = +54:18:42.10) near to the star forming H II region, NGC 5461 (see Fig. \ref{m101image}).  In line with prior studies \citep[eg][]{2023ApJ...953L..16H}, an explosion date of JD$=2460083.25$ is used throughout this paper.  Within 1 day of explosion, it was classified as a Type II SN by \citet{2023TNSAN.119....1P} based on its spectrum which showed a blue continuum with emission lines of H, He, N, and C.  High ionisation flash spectral lines were observed which are understood to result from shock interaction with a dense CSM \citep{2017NatPh..13..510Y}.

Several groups reported the results from pre-explosion space telescope photometry at the apparent SN site with a very wide spread of progenitor mass estimates.   A relatively low $8-11$M$_{\odot}$ RSG progenitor was proposed by several groups \citep[eg][]{Kilpatrick_2023, 2023ApJ...953L..14P}.  A slightly larger mass range of $12-14$M$_{\odot}$ was estimated by \citet{2024ApJ...968...27V}. Others have proposed a much larger mass range of $17-24$M$_{\odot}$ \citep{2023ApJ...952L..30J, 2023ApJ...955L..15N, 2024MNRAS.534..271Q, 2023ApJ...957...64S} potentially making this star one of the most massive SNe-II progenitors ever directly detected.  Most observations identified that the progenitor appeared to be a dusty, periodic ($\sim$1,000 day) RSG experiencing mass loss.

A complex and extended CSM was inferred from the early phase Light Curve, Spectrum and Spectropolarimetry  \citep{2023ApJ...956L...5B, 2023ApJ...955L...8H, 2024arXiv240807874H, 2023ApJ...954L..42J, 2025A&A...694A.319K, 2024Natur.627..754L,  2024PASJ...76.1050M, 2024ApJ...975..132S,  2023ApJ...956...46S, 2023ApJ...954L..12T,     2023ApJ...955L..37V, 
 2024Natur.627..759Z} suggesting the progenitor star had experienced considerable asymmetric mass loss prior to explosion.  Its early light curve appeared to show a drop consistent with that of a Type II-L SN \citep{2023TNSAN.213....1B}.  Although it showed a plateau phase of almost constant luminosity \citep{2023ApJ...954L..42J} suggesting a possible Type II-P classification.  Through hydrodynamical modelling of the light curve, \citet{2024A&A...681L..18B} derived a progenitor mass of $M_{ZAMS}=12 $M$_\odot$, an explosion energy of 1.2 $\times 10^{51}$ erg and a $^{56}$Ni mass of 0.05 M$_\odot$, making SN 2023ixf a relatively typically Type II explosion.

SN 2023ixf continues to be the subject of considerable observational focus and we have recently seen the first results from nebular phase analysis.  \citet{2024A&A...687L..20F} published the first nebula spectrum of SN 2023ixf taken at day +259 which showed broad, asymmetric emission features consistent with spectral models of a progenitor star with M$_{ZAMS} = 12-15$ M$_\odot$.  

In this paper, we focus on spectroscopic analysis of the nebular phase between day +291 and +413.  We support this with high cadence photometry to infer properties of the SN progenitor.  At the time of writing, \citet{2025MNRAS.tmp..296K} published an analysis of a nebular spectrum of SN 2023ixf on day +363 using the 4.2m William Herschel Telescope.  They identified a peculiar multi-peaked Hydrogen profile with broad, blue- and red-shifted features, potentially powered by Shock Interaction with CSM.  The nebular Oxygen lines were used to infer a relatively low progenitor mass of $M_{ZAMS}<12$M$_\odot$.  The epoch of their nebular spectrum falls between the second and third observation reported in this study and therefore, provides a useful comparison. 

This paper is organised as follows. Section \ref{methods_section} outlines the observational and data reduction methods used.   Section \ref{lightcurvesection} provides a brief analysis of the SN 2023ixf light curve to compare to spectroscopic data.  Section \ref{Specresults} outlines the results from the analysis of SN 2023ixf's nebular spectrum.  Section \ref{DiscussionSection} provides a discussion around these results while Section \ref{ConclusionSection} provides concluding remarks.   

\section{methods}
\label{methods_section}

\subsection{Photometry}
\label{Photometry}

\begin{figure*} 
\hspace*{-1.5cm}
\includegraphics[width=20cm]{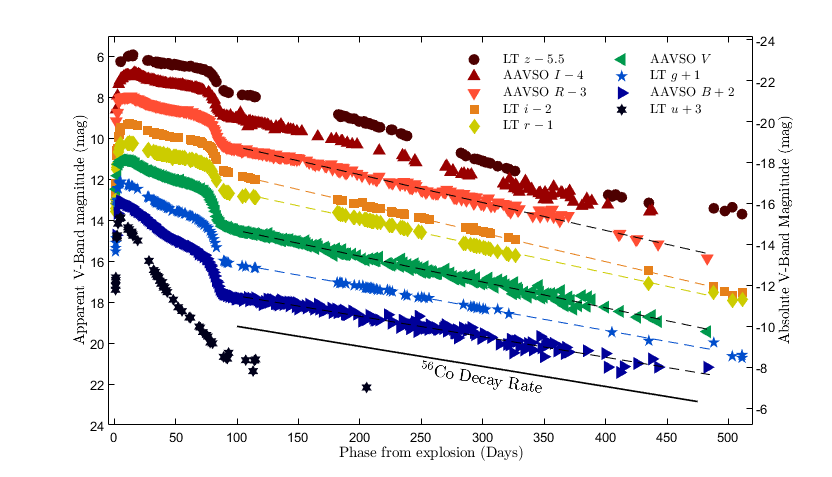}
    \caption{Multi-band light curve of SN 2023ixf from explosion to day +500.  Dashed lines show the best fit to several filter bands which were observed to have a smooth linear decline in magnitude space.  The black solid line represents the $^{56}$Co decay rate.  $BVRI$ data taken from AAVSO and $ugriz$ data taken from LT, see section \ref{Photometry}}
\label{SN2023ixfLightCurve}
\end{figure*}

Photometric data was obtained from the Liverpool Telescope (LT, \citet{2004SPIE.5489..679S}) and combined with public data contributed by The American Association of Variable Star Observers (AAVSO) to cover a wide range of filter bands.

LT Observations were taken with the IO:O optical imager in the Sloan Digital Sky Survey (SDSS) $u, g, r, i$ and $z$ filters.  Reduced images were downloaded from the LT archive and processed with custom image-subtraction and analysis software (Hinds et al., in prep.).  Image stacking and alignment was performed using \texttt{SWarp} \citep{2002ASPC..281..228B}  where required.  Image subtraction was performed using a pre-explosion reference image in the appropriate filter from the Panoramic Survey Telescope and Rapid Response System 1 (Pan-STARRS1) or SDSS.  The photometry was measured using PSF fitting methodology relative to Pan-STARRS1 (for $g, r, i$ and $z$) or SDSS (for $u$ only) standards and is based on techniques in \citet{2016A&A...593A..68F}. Although Pan-STARRS1 and SDSS covered the field, we calibrated these fields and created a catalogue of stars using IO:O standards taken on the same night at varying airmasses and used these observations to calibrate the photometry \citep[see][]{2002AJ....123.2121S}.  High cadence observations took place over the first 100 days but ceased between day +115 and day +180 due to an enclosure hydraulic fault with LT.  Regular observations then continued until day 500.

The AAVSO began a concerted photometric observing campaign of SN 2023ixf following explosion.  As of 12 September 2024, 17,222 high cadence photometric measurements have been taken from several hundred observers around the world including over 100 measurements on a number of nights near to maximum brightness.    By binning each AAVSO observation into 1 day intervals it was possible to construct relatively high-cadence $BVRI$-band photometry for the first 300 days.  After this point, the cadence of observations had become too low to perform meaningful sampling statistics although ad hoc observations continued beyond day +450 which were individually included.  

This combined photometric dataset was used to produce a $uBgVriRIz$ multi-band light curve covering the first 500 days.

\subsection{Spectroscopy}
\label{DataReduction}

3.5 hours of telescope time was obtained to use the SPectrograph for the Rapid Acquisition of Transients (SPRAT) \citep{2014SPIE.9147E..8HP} to obtain spectra of SN 2023ixf during its nebular phase in the Spring of 2024.  SPRAT is a low resolution (R=350), optical Spectrograph mounted on the LT which covers a wavelength range of $\sim4000 - 8000$\AA.

\begin{table}
\begin{center} 
\begin{tabular}{lccr} % four columns, alignment for each
\hline
JD &	Epoch  & Integration  & Observational notes\\
(UTC) & (Days) &  Time (s)  \\
\hline
2460374.7 & +291	&	1,500	&	Non-photometric, average seeing \\
2460425.6& +342	&	2,100	&	Photometric, average seeing \\
2460463.4 & +380	&	3,600	&	Non-photometric poor seeing \\
2460496.4 & +413	&	4,700	&	Photometric, good seeing \\
\end{tabular}
\caption{Nebular phase Spectroscopic observations of SN 2023ixf taken as part of this study}
\label{observations}
\end{center}
\end{table}

Four nebular spectra of SN 2023ixf were obtained between day +291 and day +413 after explosion.  The "Red Optimised" grism configuration was used to optimise the signal-to-noise ratio of the longer wavelength range since the nebular features of primary interest reside in the 6000-7500 \AA$\,$ region. One exposure of the spectrophotometric standard star, BD+33 2642 was requested immediately after each set of observations for absolute flux calibration under as close to identical conditions as possible.  This standard star was chosen due to the close proximity to M101 in the sky and a well defined flux density in the 4000-8000 \AA $\,$ range \citep[see][]{1990AJ.....99.1621O}.

A summary of the four nebular spectral observations can be seen in Table \ref{observations}.  The attempted imaging of the spectrophotometric standard failed on day +291 (so an alternative spectrophotometric standard was used) and particularly poor seeing was noted on day +380 reducing the quality of that observation.  

Spectra were reduced using a bespoke Python routine (see Section \ref{Ack} for various libraries used).  Cosmic Rays were corrected using the python package \texttt{lacosmic}  \citep{2001PASP..113.1420V}.  Differences in the airmass of the standard star and science frames were corrected for by applying Table 1 from \href{ https://www.ing.iac.es/Astronomy/observing/manuals/ps/tech_notes/tn031.pdf}{La Palma Technical Note No. 31}.  Due to the close proximity of NGC 5461 (see Fig. \ref{m101image}), a manual process was followed to carefully extract the background but several spectra risk potential contamination from this HII region.  Finally, each spectrum was scaled to ensure consistency with the observed broadband photometry and to mitigate any potential slit loss or atmospheric differences between the long-exposure science image and standard image used for calibration. 

\subsection{Nebular Spectra Sample and Features Investigated}

To draw comparisons with SN 2023ixf, a sample of CC-SNe nebular spectra was obtained from The Weizmann Interactive Supernova Data Repository \citep[WiseRep][]{2012PASP..124..668Y} corresponding to $250-450$ days after explosion.   All spectra were corrected for host galaxy redshift and extinction using the relationship of \citet{1989ApJ...345..245C} and the values in Table \ref{Tab:SNeSampleData}.  The nebular spectra of SN 2023ixf were corrected for extinction using $E(B-V)=0.039$ composed of Milky Way extinction of $E(B-V)=0.08$ \citep{2011ApJ...737..103S} and M101 host galaxy extinction of $E(B-V)=0.031$ \citep{2023TNSAN.160....1L}.

The emission lines of interest were fitted using Gaussian functions.   Strongly blended features such as the [Ca II]$\lambda\lambda7291,7324$ region required the fitting of several lines with multiple parameters simultaneously.  This was conducted following a Markov Chain Monte Carlo (MCMC) routine with 10,000,000 iterations to obtain a fit reported as the mean value of the model parameter (excluding a burn in of 25$\%$) and an error estimate using the 2$\sigma$ standard deviation of the parameter's posterior distribution.  Section \ref{OxygenAbund} and \ref{FeNiII} provides a more detailed overview of the line fitting approaches used for these more complex profiles. 

Comparison SNe flux measurements were converted to Luminosities using the distance values of Table \ref{Tab:SNeSampleData}.  With the exception of SN 1987A, uncertainty in the distance to host galaxies became the dominant source of error in Luminosity estimates.  

\section{SN 2023\lowercase{ixf} Lightcurve and $^{56}$Ni mass}
\label{lightcurvesection}

\begin{figure}
	\includegraphics[width=\columnwidth]{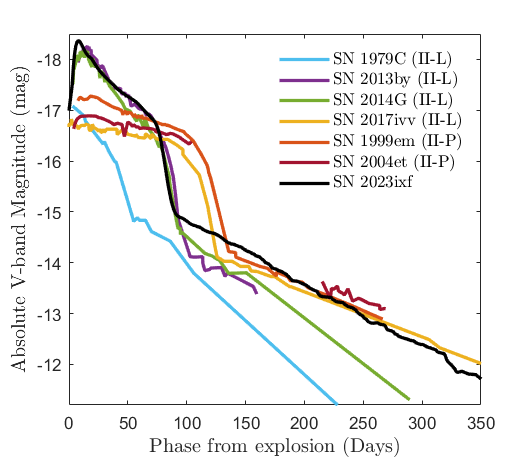}
    \caption{V-Band Lightcurve of SN 2023ixf compared to a sample of II-L and II-P SNe.  Light curves were taken from the studies referenced in Table  \ref{Tab:SNeSampleData}}
\label{lightcurvecomparison}
\end{figure}

The multiband apparent- and absolute- lightcurve (with $\mu$=29.2, see Section \ref{methods_section}) can be seen in Fig. \ref{SN2023ixfLightCurve}.  The underlying photometric data for this Figure can be found in Supplementary Data File 1.  SN 2023ixf reached a maximum brightness of $M_V=-18.2$ followed by a plateau phase which lasted until day $75\pm2$ before showing a 1.6 mag drop in brightness.  The mid-point of this drop occurs at $82 \pm 2$ days following explosion.  These estimates are consistent with several prior studies which found that SN 2023ixf had a relatively short plateau phase \citep[eg.][]{2024ApJ...975..132S, 2024A&A...681L..18B, 2024arXiv240807874H} compared to the typical 100+ day plateau for II-P SNe \citep{2014MNRAS.442..844F}; suggesting SN 2023ixf had a thinner Hydrogen envelope.  The early light curve of SN 2023ixf is characteristic of II-L SNe which have a shorter plateau duration of $80-100$ days \citep{2014MNRAS.442..844F, 2015MNRAS.448.2608V}. As noted by \citet{2024ApJ...975..132S}, the peak luminosity and the plateau length, decline rate and drop show a remarkable similarity to the fast declining II-L SNe examples SN 2013by and SN 2014G illustrated in Fig. \ref{lightcurvecomparison}.

The radioactive tail phase begins around day +90. During this nebular phase, the $B$-band decays at $1.00\pm0.03$ mag(100 days)$^{-1}$; in line with the decay rate of $^{56}$Co.  All other observed bands decayed steeper than the decay rate of $^{56}$Co suggesting an incomplete trapping of $\gamma$-rays.  Between day +150 and +300, SN 2023ixf appears to show a similar absolute V-band brightness to the prototypical examples of II-P SNe highlighted in Fig. \ref{lightcurvecomparison}.   However, with a $V$-band decay rate of $1.26\pm0.01$ mag(100 days)$^{-1}$ SN 2023ixf declines at a much steeper rate than the II-P SNe SN 1999em (0.97 mag(100 days)$^{-1}$, \citet{2003MNRAS.338..939E}) and SN 2004et (1.17 mag(100 days)$^{-1}$,\citet{10.1111/j.1365-2966.2010.16332.x}). This V-band decline is slower than the mean decline rate of 1.47 mag(100 days)$^{-1}$ observed in larger samples of Hydrogen rich SNe \citep{2014ApJ...786...67A} and slower than the fast declining II-L SNe examples highlighted in Fig \ref{lightcurvecomparison}. The II-P like nebular phase V-band brightness of SN 2023ixf may be the result of additional H-$\alpha$ emission observed at this time and discussed later in Section \ref{H-feature-section}.

The $R$ through $B$ bands all appear to be approximately linear in magnitude space between days +100 and +500 and the decay rate is generally steeper at longer wavelengths.  However, a slight upward bend is noted in the $I$ and $z$ bands after day +300.  This potential IR excess may be due to interaction with either pre-existing or newly formed dust \citep[eg][]{1985ApJ...297..719D, 2003Natur.424..285D, 2005ApJ...627L.113B}.  Unfortunately, data points are more sparse at this epoch so longer term studies in the Near-IR regime would be required to determine if this trend continues.

\begin{figure}
	\includegraphics[width=\columnwidth]{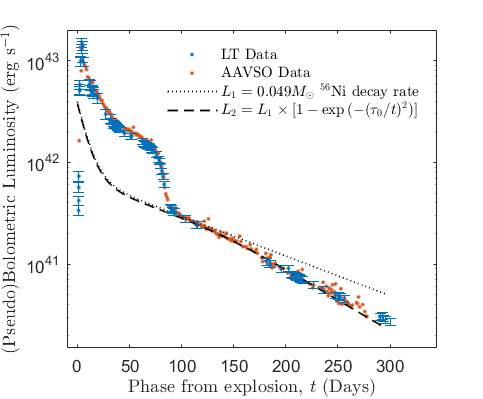}
    \caption{Bolometric light curve of SN 2023ixf derived from the independent LT and AAVSO data (see section \ref{methods_section}) using the bolometric conversion of \citet{2022A&A...660A..40M} from explosion to day +300 illustrating that the early radioactive tail phase closely tracks the luminosity output of an initial 0.049 $\pm 0.005M_\odot$ of $^{56}$Ni ($L_1$, dotted line) but decays steeper in the later phase which can be represented with an additional term accounting for the incomplete trapping of $\gamma$-rays with a timescale of $\tau_{tr}=240$ days ($L_2$, dashed line)}
\label{SN2023ixfBolometric}
\end{figure}

The AAVSO and LT photometric data were used to create two Bolometric lightcurves plotted together in Fig. \ref{SN2023ixfBolometric} using the Bolometric conversions from \citet{2022A&A...660A..40M}.  The two independent photometric datasets show good agreement and an estimate of the $^{56}$Ni mass was obtained by applying Equation \ref{eq:56Nimassat_t} \citep[see][]{1994ApJS...92..527N} which assumes that at time, $t$, the observed bolometric Luminosity, $L(t)$, is equal to the instantaneous decay of $^{56}$Co and $^{56}$Ni from an initial $^{56}$Ni mass, $M(^{56}$Ni), where

\begin{align}
        \frac{M\left(^{56}\text{Ni}\right)}{\text{M}_\odot} = \frac{L(t) / 10^{43}}{6.45 \exp{(-t/8.8)} + 1.45  \exp{(-t/111.3)}}
    \label{eq:56Nimassat_t}
\end{align}
\\

A mean value of M($^{56}$Ni$)\approx0.05 $M$_\odot$ is implied from the early tail phase light curve (days +100 to +130).  However, the later Bolometric luminosity decays significantly faster than the expected  $^{56}$Co decay rate (see dotted line in Fig \ref{SN2023ixfBolometric}) again suggesting an incomplete trapping of $\gamma$-rays.  \citet{1997ApJ...491..375C} describe a metric to account for the incomplete trapping of $\gamma$-rays within SE-SNe, where the observed Luminosity is given by:

\begin{align}
L_\text{obs}= L_{^{56}\text{Ni}} ( 1 - \exp(-(\tau_{tr} / t)^2 ))
    \label{eq:trappingadj}
\end{align}  

Where $\tau_{tr}$ is the full trapping characteristic timescale defined as:

\begin{align}
\tau_{tr} = \left( D \kappa_\gamma \frac{M^2_\text{ej}}{E_\text{K}} \right) ^{1/2}
    \label{eq:trappingtimescale}
\end{align}  

Where $D$ is the density profile constant ($\frac{4}{90 \pi}$ for a uniform density profile) and $\kappa_\gamma$ is the $\gamma$-ray opacity (estimated as 0.03 cm$^2$ g$^{-1}$ by \citet{1980ApJ...237L..81C}).

A two parameter MCMC routine combining Equations \ref{eq:56Nimassat_t} and \ref{eq:trappingadj} estimates a mean $^{56}$Ni mass of $0.049\pm0.005 $M$_\odot$ and $\tau_{tr}=240 \pm 4$ days.  This $^{56}$Ni mass estimate is consistent with a number of prior studies \citep[eg.][]{2024PASJ...76.1050M,2024ApJ...975..132S} but is significantly lower than the $0.07-0.1 $M$_\odot$ derived by others \citep[eg.][]{2024Natur.627..759Z, 2024ApJ...969..126Y}.   The variation in these $^{56}$Ni mass estimates may be due to additional sources of uncertainty which were more difficult to model here.  A Bolometric conversion was used in the present study which may underestimated the IR and UV energy contribution to Bolometric luminosity; resulting in an underestimate the $^{56}$Ni mass.  Conversely, CSM interaction during the early nebular phase may already begin to contribute additional luminosity; leading to an overestimate of the $^{56}$Ni mass.

The implied Luminosity evolution using these values is plotted as a dashed line in Fig. \ref{SN2023ixfBolometric} showing good agreement with the observed Bolometric luminosity over this period. It was not possible to investigate the Bolometric luminosity trend beyond 300 days using this method due to the scope of the conversions from \citet{2022A&A...660A..40M}.

\subsection{Explosion Energy, Ejected Mass and Progenitor Radius}
\label{explosionparameters}

To further explore the progenitor properties implied by the light curve features, the explosion energy, $E_\text{exp}$, ejected mass, $M_\text{ej}$ and the radius of the progenitor, $R_\text{prog}$ were estimated using the analytic relations of \citet{1993ApJ...414..712P}:  

\begin{align}
\log{(E_\text{exp})} &= 4.0 \log t_\text{p} + 0.4 M_V + 5.0 \log{(v_\text{ph})} -4.311 \\ 
\log{(M_\text{ej})} &= 4.0 \log t_\text{p} + 0.4 M_V + 3.0 \log{(v_\text{ph})} -2.089\\
\log{(R_\text{prog})} &= -2.0 \log t_\text{p} - 0.8 M_V - 4.0 \log{(v_\text{ph})} -4.278\\ \nonumber
\end{align} 

Where $t_p$ is the plateau length which for SN 2023ixf is measured as $82\pm 2$ days, $M_V$ is the V-band absolute magnitude in the middle of the plateau which is measured as $-17.3 \pm0.3$ mag, and $v_\text{ph}$ is the photospheric velocity at $t_\text{p}/2$ which is commonly estimated from the blue shift of the [Fe II] $\lambda$5169 absorption line at this phase.  The spectrum on day +44 (see Fig. \ref{fullspecevo}) is the closest observable to this period which shows this feature centred on 5083 \AA $\,$ corresponding to a photospheric velocity of 4,989 $\pm$ 212 kms$^{-1}$.

These observational parameters give an estimated explosion energy of 0.82 $\pm$ 0.55 $\times 10^{51}$ erg, an ejecta mass of 5.5 $\pm$ 3.0$ $ M$_\odot$ and a progenitor radius of 873 $\pm$ 443 R$_\odot$.  These estimates are in line with other published estimates \citep[eg][]{2024ApJ...975..132S, 2024ApJ...969..126Y, 2024Natur.627..759Z, 2025A&A...694A.319K} and point to a relatively typical explosion energy \citep[see][]{2009ApJ...703.2205K} and RSG progenitor radius \citep[see][]{2015PASA...32...16S}, but a relatively low ejecta mass \citep[see][]{2022A&A...660A..41M} for a Type II explosion.

Applying Equation \ref{eq:trappingtimescale} and using an energy, $E_\text{K}$ of 0.8 $\times 10^{51}$ erg, a trapping timescale of $\tau_{tr}=240 \pm 4$ days (and assuming the constants provided), gives an Ejecta Mass of 6.9$\pm$2.2 M$_\odot$.  This is again on the lower end of RSG mass estimates suggesting the progenitor has experienced some degree of mass loss to account for the shorter plateau, higher maximum brightness and steeper decline of the light curve.

\section{Spectral Evolution of SN 2023ixf}
\label{Specresults}
\begin{figure*}
\hspace*{-3cm}
	\includegraphics[width=23cm]{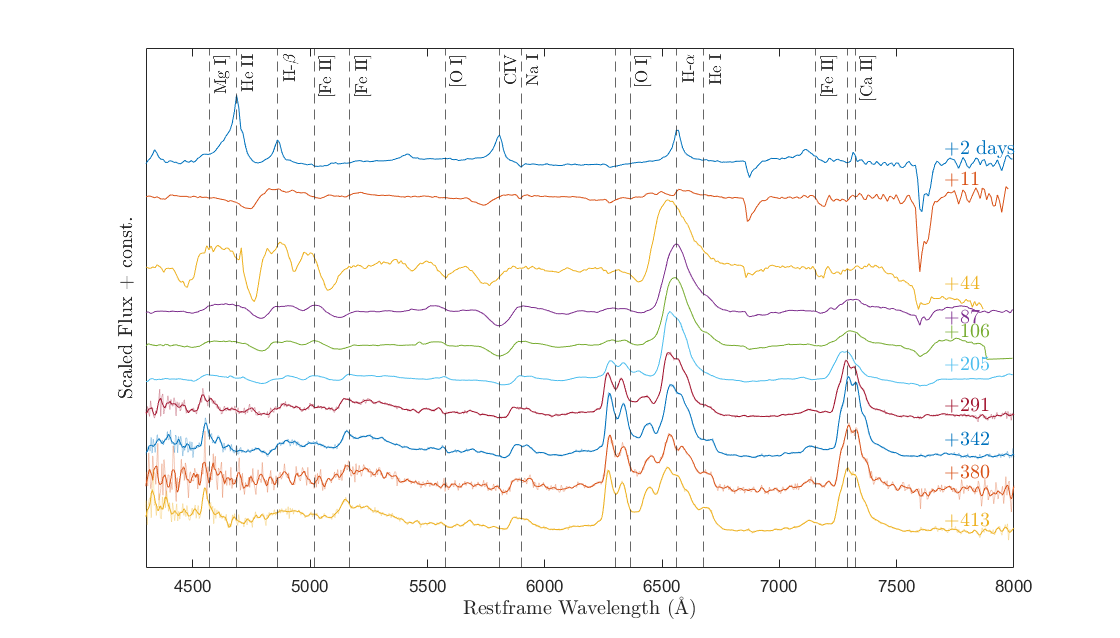}
    \caption{Spectroscopic evolution of SN 2023ixf from maximum brightness to the nebular phase, highlighting the major absorption and emission lines.  Spectra from day +2 (max brightness) to day +205 are taken from \citet{2023TNSAN.157....1P} and calibrated using the SPRAT automated pipeline \citep{2012ASPC..461..517B}.  Nebular spectra from day +291 to +413 were reduced following the method in Section \ref{methods_section} and have been overlaid with a Savitzky–Golay filter to reduce noise for clarity}
\label{fullspecevo}
\end{figure*}

\begin{figure}
	\includegraphics[width=\columnwidth]{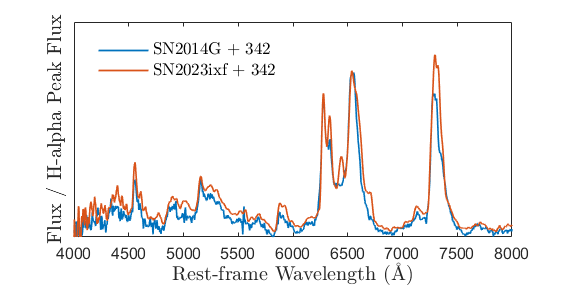}
    \caption{Comparison between the Day +342 spectra of SN 2023ixf (which has been smoothed with a Savitzky–Golay filter to reduce noise) and the II-L SN 2014G from \citet{2016MNRAS.462..137T}.  Both spectra have been normalised to their peak H-alpha flux}
\label{sn2014gComp}
\end{figure}

The spectral evolution of SN 2023ixf is shown in Fig. \ref{fullspecevo}.  The spectra presented from day +2 (max light) to day +205 after explosion were taken from the Liverpool Telescope’s Public Observing Campaign of SN 2023ixf \citep{2023TNSAN.157....1P} and are presented as they appear directly from the SPRAT automated pipeline \citep{2012ASPC..461..517B} which carries out an automated extraction, wavelength and flux calibration of spectra.  

A notable feature of the Day +2 spectra is the narrow emission lines including He II and C IV.  These highly ionised lines in the very early spectra of Type II SNe are characteristic of shock interaction with confined circumstellar material \citep{2023ApJ...954L..42J}.  After day +2, the early spectra are dominated by broad absorption lines.  From day +40, we see the emergence of emission lines and strong P-Cygni profiles. 

Forbidden nebular emission lines began to be visible by around day +87 and dominate the spectral features by day +291.    Most features are blue shifted after day +44 which is a common feature observed in II-P SNe \citep{2014MNRAS.441..671A} and has often been associated with the formation of dust as first observed in SN 1987A \citep{2016MNRAS.456.1269B}.  This, in addition to the low ejecta mass and incomplete $\gamma$-ray trapping, may account for the steep Bolometric lightcurve decline seen in Fig. \ref{SN2023ixfBolometric}.

The spectra from day +291 to day +413 were reduced following the procedure outlined in section \ref{DataReduction}.  A summary of the measurements of the main nebular emission lines is presented in Table \ref{Tab:SpecObs}.

\begin{table*}
\begin{tabular}{lccccc} % four columns, alignment for each
		\hline
		Epoch  (Days  & FWHM Velocity$^1$ & Luminosity$^2$   & Fraction of $^{56}$Ni & Blueshift$^3$ \\
  from explosion) & (km s$^{-1}$) & ($10^{38}$ erg s$^{-1})$  & Luminosity $(\%)$ & (km s$^{-1}$) \\

		\hline
            & & H-alpha  &   \\
            +291 & 6003 $\pm$ 680 & 36.2 $\pm$ 5.8 & 7.0 $\pm$ 1.1 & 295 $\pm$ 287 \\
            +342 & 6057 $\pm$ 857 & 17.3 $\pm$ 2.7 & 5.2 $\pm$ 0.8 & 162 $\pm$ 187 \\
            +380 & 7253 $\pm$ 755 & 9.9 $\pm$ 1.2 & 4.2 $\pm$ 0.5 & 290 $\pm$ 153 \\
            +413 & 8281 $\pm$ 593 & 7.6 $\pm$ 0.7 & 4.4 $\pm$ 0.4 & 632 $\pm$ 69 \\
            \hline
            & & [O I] $\lambda \lambda$ 6300,6364 \\
            +291 & 2023 $\pm$ 78 &  10.1 $\pm$ 0.62 (1.26) & 1.95$\pm$ 0.13 & 1583 $\pm$ 68 \\
            +342 & 2101  $\pm$ 133 & 6.5 $\pm$0.6 (1.2) & 1.98 $\pm$ 0.18 & 1120 $\pm$ 121\\
            +380 & 2097  $\pm$ 179 & 3.8 $\pm$ 0.4 (1.33)  & 1.65 $\pm$ 0.19 & 1136 $\pm$ 158\\
            +413 & 2086  $\pm$ 117 & 2.5$\pm$ 0.2  (1.25)& 1.44  $\pm$ 0.12 & 1402 $\pm$ 120 \\
\hline
            & & [Ca II] $\lambda\lambda7291,7324$ \\
            +291 & 2875 $\pm 196$ &  14.7 $\pm$ 1.1 (1.0)& 2.8 $\pm$ 0.2 & 649 $\pm$ 72 \\
              +342  & 2983 $\pm 150$ & 8.8 $\pm$ 0.5 (1.0)& 2.7 $\pm$ 0.2 & 66  $\pm$ 72    \\
            +380  & 2870 $\pm$ 75& 5.1 $\pm$ 0.3 (1.0)& 2.1 $\pm$ 0.1 &  136 $\pm$ 320 \\
            +413 & 3012 $\pm$ 81 &  3.0 $\pm$ 0.2 (1.0)&1.7 $\pm$ 0.1 & 318   $\pm$ 218  \\
\hline
& & [Fe II] $\lambda$7155 \\
            +291 & 2773 $\pm$ 68 & 1.16 $\pm$ 0.09  & 0.25 $\pm$ 0.07& 949 $\pm$ 108 \\
            +342 & 2993 $\pm$ 109 & 0.78 $\pm$ 0.14 & 0.24 $\pm$ 0.04 & 842 $\pm$ 84     \\
            +380 & 2893   $\pm$ 141 &0.59 $\pm$ 0.10 & 0.25 $\pm$ 0.04 & 870 $\pm$ 186  \\
            +413 & 2480 $\pm$ 74& 0.31 $\pm$ 0.06 & 0.18 $\pm$ 0.03 & 980 $\pm$ 392    \\

\end{tabular}
\caption{Summary of SN 2023ixf spectral emission features measured.  $^1$H-alpha central region and individual velocity of [O I]$\lambda \lambda$ 6300,6364 and [Ca II]$\lambda\lambda7291,7324$ doublets; $^2$ Number in brackets represents the line ratio of red to blue Gaussian features, for H-$\alpha$, luminosity represents the combined luminosity of the triple peaked emission feature; $^3$Blueshift represents the central peak of Gaussian fit}
\label{Tab:SpecObs}
\end{table*}

\subsection{Hydrogen Features}
\label{H-feature-section}
\begin{figure}
	\includegraphics[width=\columnwidth]{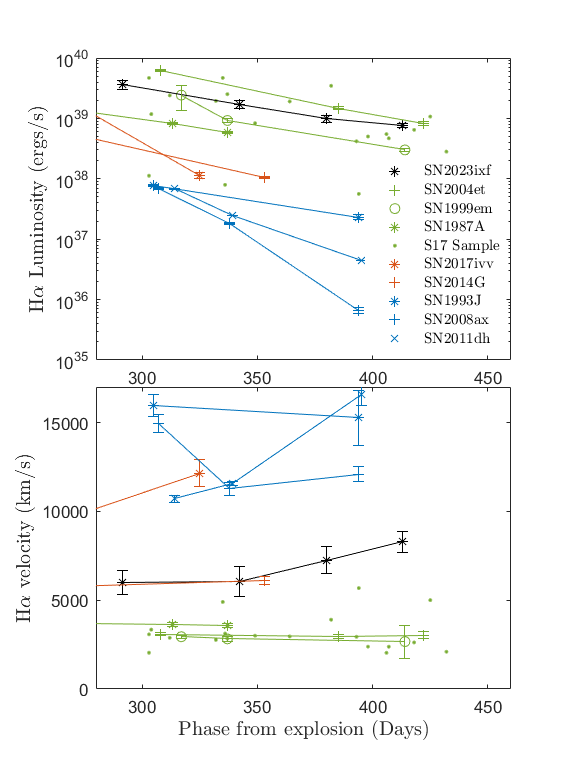}
    \caption{Total H-$\alpha$  luminosity (top) and central velocity (bottom) evolution of SN 2023ixf with a range of other CC-SNe.  Examples of IIb SNe are shown in blue, II-L SNe in red and II-P SNe in green (with the large \citetalias{Silverman2017} sample of II-P SNe shown in green dots).  SN 2023ixf showed a greater Hydrogen velocity than typical H-rich II-P SNe (green markers) but lower than examples of IIb SNe (blue markers) with significantly stripped Hydrogen envelopes. Its velocity was similar to the IIL SN 2014G.  However, its total H-$\alpha$ luminosity was more in line with Hydrogen rich examples.  See Table \ref{Tab:SNeSampleData} for references of quoted SNe}
\label{havelocity}
\end{figure}

\begin{figure*}
	\includegraphics[width=18cm]{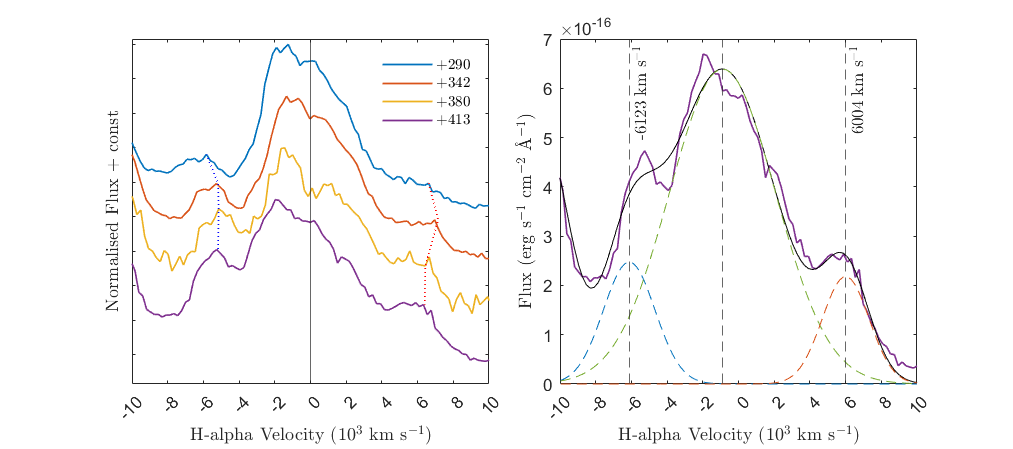}
    \caption{The nebular phase H-$\alpha$ velocity profile of SN 2023ixf.   The velocity evolution is shown on the left highlighting the peak of the blueshifted (blue dashed line) and redshifted (red dotted line) high velocity features.  The profile could be reproduced as seen on the right figure for day +413 using a triple Gaussian comprising of a very broad, central component (green dashed line) and two high velocity emissions features (blue and red dashed lines centred on approximated $\pm 6,000$kms$^{-1}$).}
\label{havelocityprofile}
\end{figure*}

SN 2023ixf showed a broad and complex Hydrogen profile throughout its spectroscopic evolution.  The FWHM of the H-$\alpha$ central velocity, comprising the majority of emission, was consistently measured over 6,000 km s$^{-1}$, consistent with the day +259 measurements from \citet{2024A&A...687L..20F}, and appeared to broaden to over 8,000  km s$^{-1}$ by day +413 (see Fig. \ref{havelocity}). 

By day +205, an unusual emission feature becomes visible between the [O I]$\lambda \lambda$ 6300,6364 and H-$\alpha$ emission lines.  It is very faint in the day +205 spectra and was not visible in any previous spectra here or in the day +141 spectrum published by \citet{2024ApJ...975..132S}.  Therefore, it appears to originate close to day +205.   The emission initially appeared so blueshifted with respect to H-$\alpha$ (7,500 km/s) that it was blended into the red shoulder of [O I].  It appears to narrow with respect to the central rest wavelength of H-$\alpha$ with time.

A highly red-shifted emission feature with respect to H-$\alpha$ then appears at around day +300.  Emission at this wavelength may be partially associated with [S II]$\lambda$6724 either from the SN or from the background of NGC 5461 which has been observed to emit forbidden Sulphur lines \citep{2024ApJ...971...87B} or possibly He I$\lambda$6678 which is present in some SE-SNe \citep{2011MNRAS.413.2140T}.  But it also appears to narrow with respect to the rest wavelength of H-$\alpha$ with time (see the red dotted lines in the left panel of Fig. \ref{havelocityprofile}) suggesting it was another high velocity Hydrogen feature in the red-shifted direction.  A similar conclusion was reached by \citet{2025MNRAS.tmp..296K} who observed these high velocity Hydrogen emission lines at day +363, falling between the second and third nebular phase observations reported here.

The nebular H-$\alpha$ region, therefore, is extremely unusual and no direct analogues exist in the literature to our knowledge.  Although a similar timing of delayed H-$\alpha$ emission was observed in SN 2014C \citep{2017ApJ...835..140M} More generally, the broad, boxy H-$\alpha$ profile is similar to examples of interaction powered SNe such as SN 1993J \citep{1995A&A...299..715P}, and more recent II-L SNe examples of SN 2017ivv \citep{2020MNRAS.499..974G} and SN 2014G \citep{2016MNRAS.462..137T}.  

The H-$\alpha$ profile could be represented by a triple Gaussian comprising one broad slightly blue-shifted central emission and two high velocity emission lines (one strongly red-shifted and one strongly blue-shifted; both by $\sim$6,000 km s$^{-1}$) throughout the nebular phase as shown in the right panel of Fig. \ref{havelocityprofile} for day +413 when the red feature was strongest.  Between days +200 and +400 the central H-$\alpha$ emission evolves from a round topped Gaussian profile to a more pointed and boxy shape.  As noted by, \citet{2025MNRAS.tmp..296K}, such a boxy profile appears to be consistent with nebular spectra models of Type II SNe whose power source is beginning to transition from radioactive decay to CSM interaction \citep{2023A&A...675A..33D}.  The lower panel of Fig. \ref{havelocity} shows the expansion velocity of the central H-$\alpha$ emission profile for SN 2023ixf compared to a  sample of CC-SNe.  SN 2023ixf had consistently broader Hydrogen features than any II-P sampled here or within the more extensive \citetalias{Silverman2017} sample of II-P SNe shown in green dots.  It appears to be more consistent with the II-L SN 2014G whose overall spectral profile similarity is highlighted in Fig. \ref{sn2014gComp}. 

\citet{2024A&A...687L..20F} and \citet{2025ApJ...978...36F} found that the nebular Hydrogen luminosity at day +259 was weak compared to model II-P SNe.  The central emission line in the period studied here is also low compared to Hydrogen rich II-P SNe.  But after integrating the total Hydrogen luminosity (including the blue and red shifted emissions), SN 2023ixf appears more in line with several II-P SNe (see top panel in Fig. \ref{havelocityprofile}), suggesting an increase in Hydrogen flux at later times.  This late time II-P similarity in Hydrogen luminosity is consistent with the photometric observations in Section \ref{lightcurvesection} which highlighted that during the nebular phase, prototypical II-P SNe had a similar brightness to SN 2023ixf in the V-band (which is expected to be dominated by [O I] and H-alpha emission during the nebular phase).

The blue-shifted high-velocity emission is notably always stronger than the red-shifted high-velocity emission.  This relative strength of the blue feature is partly due to the flux contribution from the blended [O I]$\lambda \lambda$ 6300,6364 line.  But it may also be the result of partial occultation of receding Hydrogen located on the red-shifted hemisphere; especially since the relative strength of the red high-velocity component increases with time (and therefore a decrease in opacity as the ejecta expands). 
 These unusual Hydrogen features will be discussed in more detail in section \ref{OLPM}.

\subsection{Oxygen Features}
\label{OxygenAbund}

\begin{figure}
	\includegraphics[width=\columnwidth]{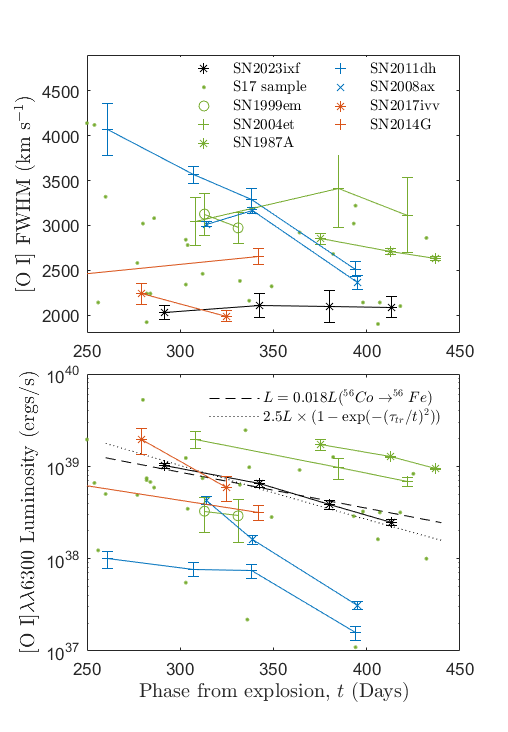}
    \caption{[O I]$\lambda \lambda$ 6300,6364 velocity (top) and luminosity (bottom) evolution of SN 2023ixf along with a sample of SE-SNe using the same colour scheme as Fig. \ref{havelocityprofile}.  The [O I] line velocity was consistently measured at the low end of SE-SNe $\sim$2,000km s$^{-1}$.  While the luminosity corresponded to 1.8$\%$ of the $^{56}$Co decay power (shown in dotted line), it was 2.5$\times$ greater at 4.5$\%$ of the $^{56}$Co decay power when accounting for the $\tau_{tr}=240$ day trapping timescale (dashed line).   See Table \ref{Tab:SNeSampleData} for references of quoted SNe.}
\label{oxygenFWHM_Luminosity}
\end{figure}

SN 2023ixf's [O I] $\lambda \lambda$ 6300,6364 emission line is narrow and double peaked throughout the nebular phase observed here.  Possible interpretations of the observed profile are discussed in Section \ref{AEF}.  But for the sake of the analysis in this section, we assume the double peaked profile represents the two individual ($\lambda$6300 and $\lambda$6364) components of the [O I] doublet. 

The Oxygen velocity was measured at FWHM<2,250 kms$^{-1}$ throughout the nebular phase which, consistent with the earlier nebular phase observation of \citet{2024A&A...687L..20F}, is one of the narrowest SE-SNe [O I] emission lines reported in the literature (see top panel in Fig. \ref{oxygenFWHM_Luminosity}).  The luminosity of [O I] is consistent with similar epoch Type II SNe sampled in this study (See lower panel in Fig. \ref{oxygenFWHM_Luminosity}).  In particular, it showed a very similar Oxygen expansion velocity and luminosity evolution to the II-L SN 2017ivv.  

Oxygen production is very sensitive to Helium-core mass and therefore the progenitor's mass \citep{1995ApJS..101..181W}.  Stellar evolution models consistently predict that more massive progenitor stars will have a larger ejected Oxygen mass \citep[eg][]{2014MNRAS.439.3694J, 1996ApJ...460..408T, 1997NuPhA.616...79N, 2003ApJ...592..404L, 2004A&A...425..649H, 2007PhR...442..269W}.  Therefore, the strength of the [O I] $\lambda\lambda$ 6300,6364 line during the nebular phase is expected to increase with progenitor mass because the more massive Oxygen shell captures an increasing fraction of the radioactive decay and because the [O I] doublet is the main coolant for the Oxygen-rich material \citep{2021A&A...652A..64D}.  But a luminosity-mass relationship under local thermodynamic equilibrium (LTE) requires an accurate measurement of the Oxygen zone temperature.  \citet{2014MNRAS.439.3694J} showed that this temperature can be constrained from the line luminosity ratio of the [O I] $\lambda$ 5577 to [O I] $\lambda \lambda$ 6300,6364 which is given by:

\begin{align}
\frac{L_{5577}}{L_{6300,6364}} =38 \exp{\left(\frac{-25,790}{T}\right)} \frac{\beta_{5577}}{\beta_{6300,6364}} 
\label{eq:oxygen_massII}
\end{align}

Where $\beta$ is the Sobolev escape probability given by: $\beta_\lambda = \frac{1 - \exp{(-\tau_\lambda)}}{\tau_\lambda}$ and $\tau_\lambda$ is the Sobolev optical depth \citep{1957SvA.....1..678S}.  Unfortunately, the assumption of LTE is only valid at the early nebular phase when the [O I] $\lambda$5577 feature is less prominent.  But even under NLTE conditions, the ratio still gives a minimum possible temperature of the [O I] zone and therefore a potentially useful maximum Oxygen mass, $M_\text{max [O I]} $ where:

\begin{align}
M_\text{max [O I]} = \frac{\left( \frac{L_{6300,6364}}{\beta_{6300,6364}} \right)}{9.7 \times 10^{41}} \times \exp{\left(\frac{22,720}{T}\right)} \text{M}_\odot 
\label{eq:oxygen_massIII}
\end{align}

\begin{figure}
	\includegraphics[width=\columnwidth]{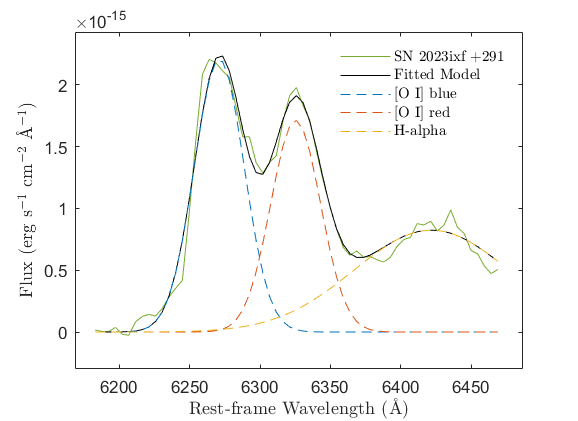}
    \caption{[O I] $\lambda \lambda$ 6300,6364 line fitting approach illustrated here with the day +291 spectrum.  The Oxygen region is represented by a double Gaussian with FWHM $= 2023\pm78$ km s$^{-1}$ and a flux ratio of 1.26.  The blue (blue dashed line) and red (red dashed line) components of the doublet are blue-shifted by 1583$\pm68$ km s$^{-1}$ relative to their resting wavelength.  A third Gaussian (yellow dashed line) was added to represent the blended high velocity H-$\alpha$ emission}
\label{OI6300LineFit}
\end{figure}

Estimating a value for $\frac{\beta_{5577}}{\beta_{6300,6364}}$ is complex but can be inferred from the observed Oxygen emission profiles.  \citet{1992ApJ...387..309L} showed that for SN1987A, a $L_{\lambda 6300}/L_{\lambda 6364}$ ratio of 1 corresponded to the early optically thick state, while a ratio of 3 is observed in the later optically thin state.   For the nebular phase observations of SN 2023ixf taken here, the [O I] $\lambda 6364$ features are always weaker than the [O I] $\lambda 6300$ with a ratio of $\sim1.20-1.33$ during the nebular phase, so the conditions are transitioning to optical thinness.  Following a similar approach to \citet{2014MNRAS.439.3694J}, this implies that the $\frac{\beta_{5577}}{\beta_{6300,6364}}$ ratio must be approximately 1.5 and within in the range $1-2$.    Their modelling work implied a range of $1.3-1.6$ for this $\beta$-ratio which we use as the uncertainty range here. Fig. \ref{OI6300LineFit} shows the line fit for the $\lambda$ 6300 region of the day +291 spectrum of SN 2023ixf.  Here, the fit corresponds to a Luminosity of $1.01\pm 0.62 \times 10^{39}$ erg s$^{-1}$.

The [O I] $\lambda$ 5577 line in the SN 2023ixf nebular spectra obtained here are very weak at most epochs.  However, a signal is detected in the day +291 spectrum.  Following a similar modelling approach to \citet{2014MNRAS.439.3694J}, the flux was estimated using a double Gaussian representing a [Fe II] line blended with the [O I] $\lambda$ 5577 emission line using the same FWHM and blueshift as measured from the stronger [O I]$\lambda \lambda$ 6300,6364 and [Fe II]$\lambda$7155 features.   The fit for the [O I] $\lambda$5577 region corresponded to a Luminosity of $1.10 \pm0.05  \times 10^{38}$ erg s$^{-1}$. 
 From equation \ref{eq:oxygen_massII}, this corresponds to a minimum Temperature of 4,122 $^{+174} _{-111}$ K (assuming $\frac{\beta_{5577}}{\beta_{6300,6364}}=1.5^{+ 0.1}_{- 0.2}$).  Finally, using equation \ref{eq:oxygen_massIII}, this corresponds to a maximum Oxygen mass of 0.51$^{+0.12}_{-0.13} $M$_\odot$ which is consistent with the measurement of 0.5 M$_\odot$ in the earlier nebular phase from \citet{2024A&A...687L..20F} and would be consistent with the Oxygen production yields of an $M_{ZAMS}<$ 15 M$_\odot$ progenitor based on a number of stellar evolution models  \citep[eg][]{2002ApJ...576..323R,2003ApJ...592..404L,2007PhR...442..269W,2016ApJ...821...38S}.  This Oxygen mass is significantly larger than the mass range of $0.07 - 0.3$M$_\odot$ estimated by \citet{2025MNRAS.tmp..296K} at day +363.  We note however, that the \citet{1986ApJ...310L..35U} method used in their study places a lower limit on the Oxygen mass.  Nevertheless, their analysis is also consistent with the conclusion of a lower mass ($M_{ZAMS}<12$M$_\odot$) progenitor from the nebular [O I] emission line.  It was not possible to repeat this method at later epochs because the [O I] $\lambda$5577 emission feature became more difficult to fit and the assumptions of LTE are weaker at later stages.  

An alternative way of investigating the progenitor mass based on the observed Oxygen Luminosity is to compare the fraction of [O I] Luminosity (as a $\%$ of total $^{56}$Co power output) to spectral models of different mass progenitors \citep[see][]{2012A&A...546A..28J}.  The [O I] $\lambda \lambda$ 6300,6364 luminosity was measured between 1.3 and 2.0 $\%$ of the $^{56}$Co total luminosity between days +291 and +413 (the dotted line in Fig. \ref{oxygenFWHM_Luminosity} represents the mean 1.8$\%$ power output for illustration) which again is consistent with the fractional power output of a 15 M$_\odot$ progenitor model from \citet{2014MNRAS.439.3694J}.  However, when accounting for incomplete trapping of $\gamma$-rays, the observed [O I] Luminosity as a fraction of $^{56}$Co decay power becomes 2.5$\times$ greater ($\sim4.5\%$ of the $^{56}$Co luminosity, see black dashed line in Fig. \ref{oxygenFWHM_Luminosity}) which could suggest a more massive progenitor.  

Overall, the [O I] Luminosity appears consistent with a lower mass ($<15$M$_\odot$) progenitor.  But due to the added complexity of incomplete $\gamma$-ray trapping, a more massive progenitor is also possible.

\subsection{Calcium Features}

The [Ca II] $\lambda\lambda7291,7324$ doublet became visible in the spectrum of SN 2023ixf on around day +100 and is one of the dominant features of the late nebular phase.   The [Ca II] features show significantly faster expansion velocities than [O I] at each epoch and showed a lower blueshift suggesting the [Ca II] emission may originate from the regions closer to the Hydrogen envelope consistent with predictions from modelling \citep{1993ApJ...405..730L, 1998ApJ...497..431K, 2012MNRAS.420.3451M}.  

Calcium abundance is dependent on the explosion energy which is not explicitly known for a SN \citep{2017hsn..book..795J}, so may to be a less direct diagnostics of progenitor properties.   Nevertheless, spectral models indicate that the [O I]/[Ca II] line flux ratio generally increases with He core mass and increases at later epochs  \citep{1989ApJ...343..323F}.  Several studies have pointed to a possible correlation connecting the [O I]/[Ca II] flux ratio to the progenitor mass \citep[eg.][]{2023ApJ...949...93F}. \citet{2011AcA....61..179E} found a mean [O I]/[Ca II] flux ratio of 0.3 for Type II SNe and that in general the [O I]/[Ca II] flux ratio increased with increasing core mass.

\citet{2024A&A...687L..20F} reported a [O I]/[Ca II] flux ratio of 0.51 on day +259.  We observed an increase in the flux ratio from $\sim0.68$ on day +291 to $\sim0.84$ at day +413 (See Table \ref{OtoCaRatio}).  Estimating this flux ratio was complicated by the fact that the [O I] emission line is blended with the high velocity H-$\alpha$ feature (see Fig. \ref{OI6300LineFit}) and the [Ca II] line includes contribution from [Fe II] and [Ni II] (see Section \ref{FeNiII}.)     Consistent with the conclusions of \citet{2024A&A...687L..20F}, this ratio appears to correspond to the expected evolution of a low He core mass of $\sim$3.1 M$_\odot$ for a $M_{ZAMS}=12$M$_\odot$ progenitor \citep{2014MNRAS.439.3694J}.  

\begin{table}
\begin{center}
\begin{tabular}{lc} % four columns, alignment for each
\hline
		Epoch  & [O I]$\lambda \lambda$ 6300,6364 to [Ca II]$\lambda\lambda7291,7324$  \\
  (Days) & Flux Ratio\\
  	\hline
291	&	0.68 $\pm$  0.10 \\
342	&	0.73 $\pm$  0.11 \\
380	&	0.75  $\pm$  0.11 \\
413	&  0.84 $\pm$  0.12 \\

\end{tabular}
     \caption{Line flux ratios of [O I]$\lambda \lambda$ 6300,6364 to [Ca II]$\lambda\lambda7291,7324$ were measured between 0.5 and 1 at all nebular phases, consistent with models of lower mass progenitor stars}
\label{OtoCaRatio}
\end{center}
\end{table}

Although the [O I]/[Ca II] ratio is an attractive parameter because it does not depend on distance and absolute flux calibration, a number of studies have questioned the validity of this metric \citep{2022MNRAS.514.5686P} so some caution is required when basing conclusions on this ratio.

\subsection{Iron and Stable Nickel Abundances}
\label{FeNiII}

\begin{figure*}
\begin{center}
\includegraphics[width=15cm]{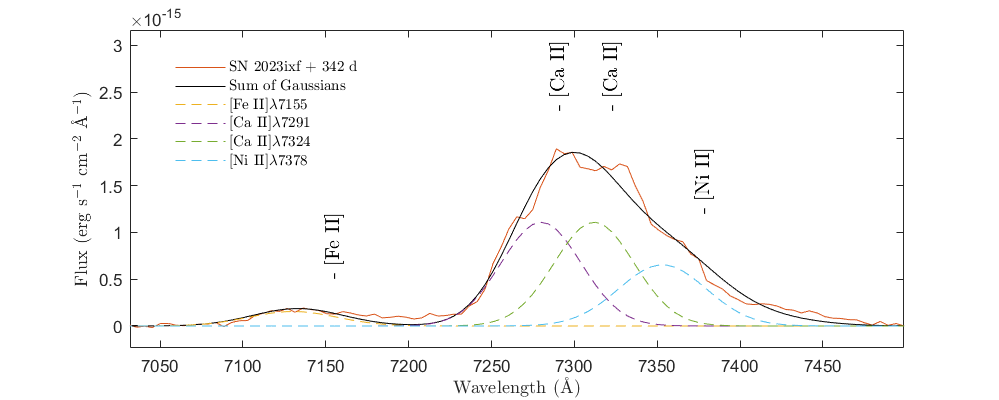}
\caption{Line fit of the [Ca II]$\lambda\lambda7291,7324$ region (which also includes contribution from several [Fe II] and [Ni II] emission lines) for SN 2023ixf at day +342 represented by eight Gaussian curves with the four strongest lines presented for clarity.  This line fit suggests the extended red shoulder of the [Ca II] emission may be partly due to [Ni II]$\lambda$7378 emission from stable Nickel}
\label{FeNiFitd290}
\end{center}
\end{figure*}

Fig. \ref{sn2014gComp} illustrated that the spectral profile of SN 2023ixf was remarkably similar to the Type II-L SN 2014G.  In particular, a similar extended red shoulder of the [Ca II]$\lambda \lambda7300$ doublet is observed.  In the case of SN 2014G, this was attributed to a high stable Nickel abundance \citep{2016MNRAS.462..137T}.   This [Ca II] region is blended with several [Fe II] and [Ni II] lines.  Here the [Ni II] $\lambda$7378 line is expected to originate from stable $^{58}$Ni, which is produced from explosive burning and can be useful in constraining properties of the explosion \citep{2015ApJ...807..110J}.  This region is shown for SN 2023ixf at day +342 in Fig. \ref{FeNiFitd290} where it has been fitted with 8 Gaussians centred close to their emission wavelength following the procedure outlined in \citet{2015MNRAS.448.2482J}.

Following the approach of \citet{2015MNRAS.448.2482J}, the ratio of line luminosities arising from the same levels were fixed such that  L$_{7172}$ = 0.24 L$_{7155}$, L$_{7388}$ = 0.19 L$_{7155}$, L$_{7412}$ = 0.31 L$_{7378}$ and L$_{7453}$ = 0.31 L$_{7155}$.   This leaves 7 free parameters to adjust in the line fit: L$_{7155}$, L$_{7291, 7323}$, L$_{7378}$, and the blue-shift and expansion velocities of [Ca II] and [Fe II],[Ni II] lines (two different velocities are used since they are expected to originate from different regions of the ejecta).  From \citet{2015MNRAS.448.2482J}, under LTE, the luminosity of the [Ni II]$\lambda$ 7378 to [Fe II]$\lambda$ 7155 is given by:

\begin{align}
\frac{L_{7378}}{L_{7155}} = 4.9 \left( \frac{n_\text{[Ni II]}}{n_\text{[Fe II]}} \right) \exp \left(  \frac{0.28 \text{eV}}{k T}\right)
\label{eq:Ni_FeRatio}
\end{align}

and the Temperature can be constrained by the Luminosity of the [Fe II]$\lambda$7155 line:

\begin{align}
\frac{L_{7155}}{M(^{56}Ni)} = \frac{8.67 \times 10^{43}}{15+0.006 T} \text{exp} \left(  \frac{-1.96\text{eV}}{kT}\right)
\label{eq:FeTemp}
\end{align}

Applying equations \ref{eq:Ni_FeRatio} and \ref{eq:FeTemp} (and the $^{56}$Ni mass obtained in Section \ref{lightcurvesection}) to the four nebula spectra taken in this study for SN 2023ixf, we obtain the temperatures and luminosities shown in Table \ref{NiFeAbundance}.   These values correspond to a mean Ni/Fe abundance of  $\sim$0.11 which is approximately 2 times the solar level of 0.056 \citep{2003ApJ...591.1220L}. 

A supersolar Ni/Fe abundance ratio has been observed in a number of CC-SNe \citep[eg.][]{2007ApJ...661..892M, 2015MNRAS.448.2482J, 2020MNRAS.499..974G, 2022ApJ...930...34T, 2024ApJ...968L..18T} and may be achieved through several possible mechanisms.   There is some evidence from modelling that lower mass progenitors are expected to have larger Ni/Fe abundance ratios \citep{1995ApJS..101..181W, 1996ApJ...460..408T}.  More recently, \citet{2015ApJ...807..110J} showed that similar Ni/Fe abundance ratios to those observed here can be achieved through either a high metallicity progenitor (over five times Solar levels) or a low mass progenitor (<13 M$_\odot$) exploding with a short delay time to eject the silicon layer with a significant neutron excess.  Studies of the explosion site of SN 2023ixf estimate this was likely to be a relatively low-to-solar metallicity environment \citep[eg.][]{2023ApJ...955L..15N, 2024ApJ...968...27V, 2024Natur.627..759Z} so this observation may be more consistent with the assumption of a lower mass explosion, providing another data-point in support of a potentially low mass progenitor.

\begin{table*}
\begin{center}
\begin{tabular}{lccccc} % four columns, alignment for each
\hline
		Epoch  & Luminosity of [Fe II]$\lambda$7155 & Luminosity of [Ni II]$\lambda$7378 & Inferred Temperature &  [Ni/Fe] &  Abundance $\times$  \\
  (Days) & (10$^{38}$erg s $^{-1})$&(10$^{38}$erg s $^{-1})$& (K)& Abundance &[Ni/Fe]$_\odot$ \\
  	\hline
291	&	1.16 $\pm$ 0.09 & 3.74 $\pm0.68$ &  2405 $\pm$ 35  &  0.18 $\pm$ 0.05 & 3.0 $\pm$ 0.9 \\
342	&	0.78 $\pm$ 0.13 & 1.67 $\pm$ 0.32 & 2300 $\pm$ 40 &  0.12 $\pm$ 0.05 & 1.9 $\pm$ 0.8 \\
380	&	0.59 $\pm$ 0.10 & 1.20 $\pm$ 0.15 & 2240 $\pm$ 50 &  0.11 $\pm$ 0.04  & 1.7 $\pm$ 0.4 \\
413	&   0.31 $\pm$ 0.06 & 0.58 $\pm$ 0.75 & 2100 $\pm$ 50 &  0.08 $\pm$ 0.04 & 1.4 $\pm$ 0.5\\

\end{tabular}
     \caption{Luminosity of [Fe II]$\lambda$7155 and [Ni II]$\lambda$7378 features, along with inferred temperature and derived Ni/Fe abundance ratios.  The multiple of solar abundance levels is shown in  column 6}
\label{NiFeAbundance}
\end{center}
\end{table*}

\section{Discussion}

We have analysed the nebular spectra of the nearby Type II SN 2023ixf and have compared its observed features to a sample of CC-SNe.   The light curve of SN 2023ixf showed a short plateau phase indicating a lower mass envelope compared to a typical II-P SN.   It also showed a very different spectral evolution to any II-P SN sampled here and in wider studies \citepalias[eg.][]{Silverman2017}.  The broad Hydrogen features in particular, appear similar to several examples of Type II-L SNe.  
 
 In this section, we will discuss what we can infer about the progenitor of SN 2023ixf and discuss possible scenarios that may explain its observed features.

\label{DiscussionSection}

\subsection{Abundances and Progenitor Mass}
\label{OLPM}

The nebular spectra of SN 2023ixf contained a number of key features which are important in revealing information about the possible progenitor of this SN.  We have shown that several analytic methods point to a relatively low mass progenitor of SN 2023ixf.  Analysis of the Oxygen line at day +291 was consistent with an Oxygen mass of <0.64 M$_\odot$ corresponding to models of a $M_{ZAMS}<15$ M$_\odot$ progenitor.  Likewise, the luminosity at all measured nebular periods was consistent with $\sim1.8\%$ of the $^{56}$Co mass power output which appears to be consistent with II-P spectral models for progenitor stars of $<15$ M$_\odot$.  The [O I] to [Ca II] flux ratio observed through the nebular phase was consistent with a low mass ($< 12$ M$_\odot$) progenitor.  Finally, the supersolar Ni/Fe abundance ratio may be explained through the explosion of a lower mass (< 13M$_\odot$) progenitor, ejecting its silicon shell.

An alternative method is available to investigate the progenitor by comparing the observed flux level to existing spectral models of II-P SNe under a range of explosion scenarios.   The model spectra are dependent on epoch, initial $^{56}$Ni mass and distance.  So for comparison, the model ($\text{mod}$) parameters can be scaled to the observed ($\text{obs}$) flux, $F$ using a distance, $d$, of 6.85 Mpc, a $^{56}$Ni mass, $M(^{56}Ni)$, of 0.049 M$_\odot$ and the time, $t$ of observation using Equation 2 from \citet{2019MNRAS.485.5120B}:

\begin{align}
\frac{F_\text{obs}}{F_\text{mod}}=\frac{d^2_\text{mod}}{d^2_\text{obs}} \frac{M(^{56}Ni)_\text{obs}}{M(^{56}Ni)_\text{mod}} \exp \left(\frac{t_\text{mod} - t_\text{obs}}{111.3} \right)
    \label{eq:modelscale}
\end{align}

\begin{figure*}
\includegraphics[width=17cm]{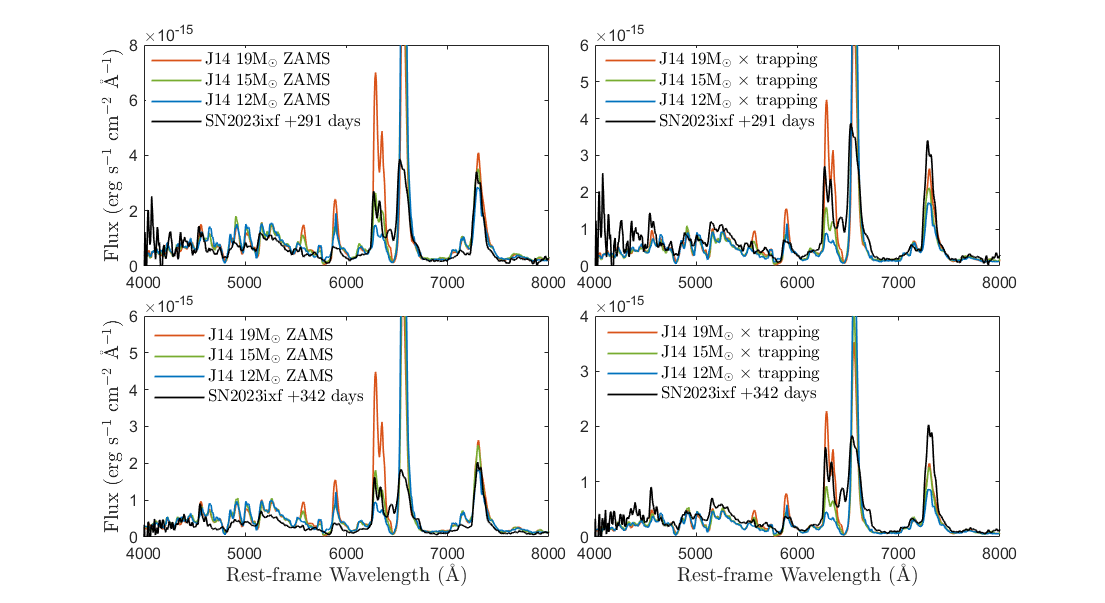}
    \caption{A comparison between the nebular spectra of SN 2023ixf (which have been smoothed with a Savitzky–Golay filter to reduce noise) and the 12M$_\odot$, 15M$_\odot$ and 19M$_\odot$ nebular spectral models of \citet{2014MNRAS.439.3694J} (J14).  The plots on the left show the day +291 and +342 Spectra of SN 2023ixf compared to the J14 models scaled using Equation \ref{eq:modelscale}.   The plots on the right show the day +291 and +342 Spectra of SN 2023ixf compared to the J14 models scaled using the additional term in Equation \ref{eq:trappingadj} to account for the incomplete trapping of $\gamma$-rays.}
\label{JerkstrandModelComp}
\end{figure*}

\citet{2014MNRAS.439.3694J} presented a number of II-P Spectra models covering days +212 to +450 for 12, 15, 19 and 25 M$_\odot$ progenitors.  The comparison between SN 2023ixf and their 12, 15 and 19 M$_\odot$ models is outlined in Fig. \ref{JerkstrandModelComp}.  Ignoring the H-$\alpha$ line which is clearly unusual in the case of SN 2023ixf, the observed spectra of SN 2023ixf correspond reasonably well to the 15M$_\odot$ model spectra at earlier epochs (see left two plots in Fig. \ref{JerkstrandModelComp}).   The 15M$_\odot$ model assumes an Oxygen mass of 0.82M$_\odot$.  This is larger than the 0.65M$_\odot$ upper limit we obtained from analysis of the day +291 [O I] $\lambda \lambda$ 6300,6364 line in this study and may therefore point to a slightly larger Oxygen mass. A similar process was followed in the earlier nebular phase by \citet{2024A&A...687L..20F} which showed a similarity to the $12-15$M$_\odot$ model spectra.  \citet{2025MNRAS.tmp..296K} found the day +363 spectra to be fainter than even the lowest mass ($M_{ZAMS}=12$M$_\odot$) model but we note their comparison assumed a $^{56}$Ni mass $\sim50\%$ greater than used here.  

A weakness of comparing these II-P SN models to SN 2023ixf is that the models assume thick Hydrogen envelopes, which does not appear to be the case here.  SN 2023ixf also had a short plateau phase (see Section \ref{lightcurvesection}) and so its brightness began decaying earlier and steeper than for a typical II-P SN.  As a result, the SN 2023ixf spectra become much fainter than the \citet{2014MNRAS.439.3694J} model spectra at later times and there is even a significant impact at the earlier two nebular spectra obtained here.   The two plots on the right of Fig \ref{JerkstrandModelComp} show the day +291 and +342 spectra of SN 2023ixf compared to the model spectra of \citet{2014MNRAS.439.3694J} adjusted to account for the different parameters in Equation \ref{eq:modelscale} and additionally Equation \ref{eq:trappingadj} in an attempt to address the difference in $\gamma$-ray trapping.  This might suggest the nebular spectra of SN 2023ixf are consistent with a more massive $15-19$M$_\odot$ progenitor.  However, the radiative transfer effects of incomplete $\gamma$-ray trapping are much more complex than highlighted here so caution must be applied to this approach.  

Overall, the nebular spectra of SN 2023ixf appears consistent with a progenitor mass of $12-15$  M$_\odot$ and an explosion energy of 1 FOE which is powered by an initial $^{56}$Ni mass of $\sim$0.049 M$_\odot$.  However, we cannot exclude the possibility of a much larger mass progenitor where the observed nebular features appear fainter at each epoch due to the incomplete trapping of $\gamma$-rays and rapidly declining light-curve. 

\subsection{Asymmetric Emission Features}
\label{AEF}
Asymmetric emission features appear common to core collapse SNe \citep{2001ApJ...559.1047M, 2012IAUS..279..261M, 2024NatAs...8..111F} and have been reported in a number of prior studies of SN 2023ixf \citep{2023ApJ...956...46S, 2023ApJ...956L...5B, 2024A&A...687L..20F}.  This study has illustrated a number of notable asymmetric emission features including an unusually high velocity, multi-peaked Hydrogen profile, a very broad Calcium emission line and very narrow, double peaked Oxygen features.

The `horn'-like [O I]$\lambda \lambda$ 6300,6364 profile has been observed in several SE-SNe \citep[eg][]{2005Sci...308.1284M, 2008ApJ...687L...9M} and has often been interpreted as emission originating from an expanding Oxygen rich torus viewed on axis resulting in a blue- and a partially occulted red-shifted emission component centred on $\sim$6300\AA$\,$ \citep[see][]{2008Sci...319.1220M, 2024NatAs...8..111F}.  This profile was present in the earlier nebular phase of SN 2023ixf and was interpreted in this way by \citet{2024A&A...687L..20F} and \citet{2025ApJ...978...36F} as evidence of an asymmetric explosion with a central trough located at $\sim$6300\AA$\,$ corresponding to the centre of the explosion (or 0 km s$^{-1}$ in velocity space centred at the blue component's rest wavelength).   However, as seen in Fig. \ref{OI6300LineFit}, the feature is also well reproduced in all nebular spectra here by a double Gaussian separated by $62 \pm 5$  \AA, consistent with the doublet's individual component's expected separation wavelength.  The flux ratio of these individual components is $\sim$1.3 (and increasing with time) which is consistent with what would be expected from the line ratio transitioning to optical thinness \citep{1992ApJ...387..309L} during this period.  The observed blueshift of the individual components is also consistent with the less strong [O I] $\lambda$5577 emission line.  So an interpretation where this profile represents emission from the narrow-line regime of the individual doublet components as suggested by \citet{2010ApJ...709.1343M} is preferred here.  While the underlying dynamics causing the observed profile is subject to debate, either interpretation does not impact the velocity or luminosity measurements reported here.

The Hydrogen profile also showed several unusual features.  By around day +200 a high velocity blue-shifted Hydrogen emission feature becomes visible.   \citet{2023ApJ...954L..12T} identified high a velocity  "Cachito" feature in the early spectroscopic period of SN 2023ixf suggesting the presence of dense Hydrogen surrounding the SN.   Such asymmetric Hydrogen profiles have been observed in a number of Type II SNe.  \citet{2017ApJ...850...89G} found that a "Cachito" emission feature between [O I] $\lambda \lambda 6300,6364$ and H-$\alpha$ was present at some stage in 60$\%$ of the 122 Type II SNe spectra observed in their sample.  In the early stages they associated it with Si II $\lambda$ 6355 and in later stages with high velocity H-$\alpha$.  A similar feature was observed in SN 1987A.  This "Bochum event" appeared to be the result of a high velocity $^{56}$Ni clump powering excess flux in the direction of the observer \citep{1988PASA....7..446H, 2002ApJ...579..671W}.  In this case the blue and red-shifted emission components narrow with respect to the H-$\alpha$ rest velocity with time confirming that they were associated with high velocity Hydrogen.   

The central expansion velocity of Hydrogen was also very broad,  appearing most similar to the II-L SNe 2017ivv and 2014G.  The Hydrogen velocity of SN 2023ixf was approximately twice that of a typical II-P SN.  Assuming similar explosion energies, this would imply SN 2023ixf has approximately 1/4 the Hydrogen mass (or $\sim2.25$M$_\odot$) of a typical II-P Hydrogen envelope ($\sim9$M$_\odot$, \citet{2012A&A...546A..28J}).

To explain this feature we need to account for several observations: 
\begin{itemize}
    \item the central Hydrogen emission shows a very high and increasing expansion velocity,
    \item the light curve analysis pointed to an average radius RSG 
    \item the Hydrogen luminosity appears greater than might be expected of a highly stripped envelope,
    \item the early lightcurve showed a plateau phase suggesting Hydrogen ionisation at a confined radius while the late-time V-band light curve resembled examples of more Hydrogen rich II-P SNe,
    \item the high velocity red and blue-shifted Hydrogen emission features only become visible after $\sim$ day +200
\end{itemize}  

These combined observations suggest that SN 2023ixf has partially retained some of its Hydrogen envelope and is showing signs of further Hydrogen emission as the SN evolves.

Since the high velocity blue shifted emission feature is first seen at day +200 (see Fig. \ref{fullspecevo}) with a velocity of 7,300 km s$^{-1}$ and assuming an ejecta velocity of $\sim$8,500 km s$^{-1}$ \citep{2023ApJ...954L..42J}, the emergence of this feature could be consistent with the SN ejecta causing shock interaction with Hydrogen-rich material at an unusually large distance of $\sim1.5 \times 10^{16}$cm ($\sim$1000 AU).  Shock interaction is further supported by the detection of a radio wave peak at day +206 by \citet{2025ApJ...978..138I} which coincides with the timing of the high velocity Hydrogen features detected here.  

Within the literature, \citet{2023ApJ...956...46S} found evidence of a dense, asymmetric, Hydrogen rich CSM surrounding SN 2023ixf.  Light curve modelling from \citet{2024ApJ...975..132S} was consistent with a two zone CSM structure surrounding SN 2023ixf: a confined dense region upto $5 \times 10^{14}$cm \citep[see also][]{2023ApJ...954L..12T} and an extended CSM spanning to at least 10$^{16}$cm.   Shock Interaction of Hydrogen within this extended CSM would be consistent with the broadening of Hydrogen emission observed in the nebular phase here.  However, while the nebular spectrum may suggest interaction power, it is worth noting that the $R$-band light curve was observed to be linear throughout the entire nebular phase (see Fig. \ref{SN2023ixfLightCurve}). It showed no signs of a knee shape which \citet{2023A&A...675A..33D} suggested would be indicative of a SN transitioning from decay to interaction power.  Therefore, during the nebular phase observed here, while CSM interaction is becoming increasingly important, the dominant power source still appears to be from the radioactive decay of $^{56}$Co.

The features of SN 2023ixf therefore, suggest four distinct Hydrogen regions surrounding the progenitor.  1) a lower mass, partially stripped Hydrogen envelope, 2) a confined region of Hydrogen rich CSM, 3) a region of lower density CSM extending to $\sim10^{16}$ cm; and 4) a dense Hydrogen rich region located at a wide radius of $\sim 1.5\times10^{16}$ cm.

\subsection{Mass Loss and the Diversity of Type II SNe}

If indeed, SN 2023ixf had a <15 M$_\odot$ progenitor, what mechanism could account for the mass loss and unusual CSM structure observed here?  RSG mass loss is still poorly understood, but broadly two main methods of Hydrogen envelope stripping have been proposed in the literature: mass loss through stellar winds or interaction with binary companions.  

Pulsation driven stellar winds can lead to significant mass loss for even a relativevly low mass RSG progenitor \citep{2010ApJ...717L..62Y}.  With stellar wind mass loss rates of upto $\sim10^{-5} $M$_\odot$year$^{-1}$ \citep{2024A&A...686A..88A}, it may appear possible for an RSG to shed $\sim4$M$_\odot$ of mass during their lifetime through this mechanism.  Assuming a wind velocity of 10 km s$^{-1}$ \citep{2023ApJ...956L...5B}, the existence of Hydrogen rich material at 1000AU would only require 1,000 years.  However, stars upto 25 M$_\odot$ are generally expected to explode as RSGs with almost all of their Hydrogen envelopes \citep{2009MNRAS.395.1409S} and models of $\sim 16$ M$_\odot$ progenitors indicate they would shed only 0.3 to 1.5 M$_\odot$ of their mass during its RSG phase \citep{2018MNRAS.475...55B}.  Stellar wind mass loss is also expected to increase with progenitor mass \citep{2014ARA&A..52..487S} but SN 2023ixf appears to have a progenitor with a very similar mass to other II-P SNe with a much greater extent of their Hydrogen envelopes intact.  Even if stellar winds could account for some of the mass loss, they may struggle to explain the unique observation of such dense Hydrogen rich material located at $>10^{16}$cm.  

Alternatively, mass loss could be due to interaction with a binary companion \citep[eg][]{1996IAUS..165..119N}.   The progenitors of CC-SNe are understood to have spent most of their lives as B- and O- type stars \citep{2019A&A...631A...5Z}.  Several studies have shown that a large proportion of these stars are located in binary systems \citep[eg][]{2012MNRAS.424.1925C} and over 70\% of massive stars are expected to exchange mass with a companion in their lifetime \citep{2012Sci...337..444S}.  In the case of SN 2023ixf, no binary companion has been identified.  Although, studies have suggested a binary companion may be possible based on pre-explosion photometry \citep{Kilpatrick_2023, 2024SCPMA..6719514X}.  Several groups have pointed to the existence of a strong asymmetric CSM surrounding SN 2023ixf as possible evidence of Roche lobe Overflow (RLOF) from a binary companion \citep{2024SCPMA..6719514X, 2024A&A...683A.154M, 2024arXiv240807874H} and this is an intriguing explanation for the unusual features observed here.  

Massive stars experiencing Case A and Case B RLOF will evolve into yellow or blue supergiants \citep{2024ARA&A..62...21M} which, as noted by \citet{2024arXiv240807874H}, is inconsistent with the RSG progenitor of SN 2023ixf observed from pre-explosion photometry \citep{2023ApJ...955L..15N}.  However, at a more evolved stage, a greater mass accretion is required to evolve stars to BSGs \citep{2024A&A...686A..45S}.  So Case C RLOF, where partial mass transfer occurs after the depletion of core He, may be a better explanation for the partial mass loss and RSG progenitor observed here for SN 2023ixf.  

A sudden swelling of CC-SNe progenitors may trigger asymmetric mass loss in binary systems \citep{2013arXiv1302.5037S, 2014ApJ...785...82S} which may explain the mass loss and asymmetric features observed here.   \citet{2024A&A...685A..58E} found that in models of wide massive stars, RLOF often led to incomplete stripping of the envelope; similar to what has been discussed here for SN 2023ixf.  \citet{2024ApJ...963..105M} found that binary interaction can account for significant diversity of CSM structure surrounding Type II SNe.  Consistent with the progenitor mass inferred here, they also found that 15 and 16.2 M$_\odot$ stars in binary systems can result in CSM structures within a scale of $\sim10^{17}$cm.  In particular, their binary models of 1300 and 1500 day periods with wind velocities of 10-100 kms$^{-1}$ led to the formation of dense shells and "cliff like" structures at length scales of $10^{16}-10^{18}$ cm, providing a possible mechanism to account for the dense Hydrogen region proposed here. 

At this stage, binary interaction is a speculative explanation for the features observed here.  However, SN 2023ixf adds to a growing list of Type II SNe showing an extraordinarily diverse range of Hydrogen features but which appear to originate from very similar low to mid-teen M$_\odot$ mass RSG progenitors.  It is difficult to explain such a diversity through single star evolutionary models alone and may potentially be better explained through external processes such as binary interaction.  

\section{Conclusions}
\label{ConclusionSection}

This analysis of the nearby Type II SN 2023ixf demonstrates that nebular phase spectroscopy can be a a useful tool for examining the inner core left over from a SN explosion.  We have used a range of analytical methods to show that the progenitor of SN 2023ixf is consistent with that of a relatively low mass Type II SN RSG progenitor (with $M_{ZAMS}=12-15$ M$_\odot$) exploding with an energy of  $\sim 1 \times 10^{51}$ erg and powered by an initial 0.049 M$_\odot$ of $^{56}$Ni.  

High Hydrogen and Calcium expansion velocities and the emergence of extended Hydrogen features suggest that the progenitor had experienced mass loss from its outer envelope.  While stellar winds may provide a general mechanism for mass loss, they may struggle to explain such a large mass loss ($\sim$6 M$_\odot$) from a relatively low mass progenitor and the presence of dense Hydrogen at an extended distance in this case.  More generally, it is also hard to reconcile the diversity of mass loss seen across Type II SNe with the similarities of their progenitors under a single star mechanism.  Therefore, mass loss via binary interaction may better account for some of this observational diversity.  In the present example of SN 2023ixf, this mass loss may potentially be consistent with Case C RLOF; implying that its progenitor star experienced significant mass transfer with a binary companion.

While nebular phase spectroscopy can be powerful for examining the progenitor of SNe, many methods discussed here rely heavily on accurate flux calibration, extinction and distance estimates.  Uncertainties in these methods and derived values can add considerable error to the conclusions presented here.  Only a very small number of nebular spectra of SN 2023ixf have been published to date, so there is limited other work in which to compare the findings presented here.  Only four nebular spectra were obtained due to the exposure times required to obtain a useful signal-to-noise ratio.  So conclusions have been based on a relatively small sample of observations.   It will be interesting to compare these results to more extended studies beyond day 500 to further explore the extended CSM structure of SN 2023ixf.

We have briefly discussed some of the theoretical binary models that may account for the unusual Hydrogen rich CSM surrounding SN 2023ixf.   Hopefully this work and other studies on the nebular phase of SN 2023ixf can be used to further test these theoretical models.  In particular, the effect of shock interaction can be further studied at Radio, X-Ray and UV wavelengths.  SN 2023ixf also offers an opportunity to investigate the effect of dust formation at IR wavelengths, in particular with JWST.  

SN 2023ixf adds to the increasing diversity of Type II SNe that appear to come from very similar progenitors.   Given the prominence of SN 2023ixf, it should be possible to study this explosion deep into the nebular period for years to come.  This may reveal further information about its unique history and enrich our understanding of the potential role of binary interaction in the diversity of observed SNe.

\section{Acknowledgements}
\label{Ack}
The Liverpool Telescope is operated on the island of La Palma by Liverpool John Moores University in the Spanish Observatorio del Roque de los Muchachos of the Instituto de Astrofisica de Canarias with financial support from the UK Science and Technology Facilities Council.  We acknowledge with thanks the variable star observations from the AAVSO International Database contributed by observers worldwide and used in this research.  In particular, extremely prompt and regular photometric data provided by Ian Sharp, which was invaluable in estimating the exposure times necessary to optimise the Signal-to-Noise ratio for the nebular spectra obtained in this research.  

Software used:  \texttt{Astropy} \citep{2022ApJ...935..167A}, \texttt{NumPy} \citep{5725236},  \texttt{SciPy} \citep{2020NatMe..17..261V}, MATLAB \citep{MATLAB:2023}.

\section{Data Availability}

Data can be shared upon request to the corresponding author.  Early phase spectral and photometric image data are available on the \href{https://telescope.ljmu.ac.uk/cgi-bin/lt_search}{LT Data Archive} under Proposal ID PQ23A01 \citep{2023TNSAN.157....1P}.  Photometric data is available in Supplementary Data File 1.  Nebular phase spectra will be made available on \href{https://www.wiserep.org/}{WiseREP} shortly after publication. 

%%%%%%%%%%%%%%%%%%%% REFERENCES %%%%%%%%%%%%%%%%%%

% The best way to enter references is to use BibTeX:

\bibliographystyle{mnras}
\bibliography{main}

\appendix
\section{Comparative SE-SNe Sample}

\begin{table*}
\setcounter{table}{0}
\caption{Sample of nebular phase CC-SNe used for comparison in this study.  All spectra were obtained from WiseRep \citep{2012PASP..124..668Y}.\\ Notes: (1) Taken from NASA/IPAC Extragalactic Database mean distance or referenced study  (2)  \citet{2022ApJ...934L...7R}  (3) \citet{2021A&A...655A.105S}
 (4) \citet{2018PASJ...70..111K} (5) \citet{2021NatAs...5..903H} (6) \citet{2020MNRAS.498...84Z} 
(7) \citet{2016MNRAS.461.2003Y} 
(8) \citet{2013MNRAS.433.1871B}
(9) \citet{2006MNRAS.372.1315S}  (10) \citet{2014MNRAS.442..844F} 
(11) \citet{10.1111/j.1365-2966.2010.16332.x}
(12) \citet{2003MNRAS.338..939E} (13) \citet{1998ApJ...500..525S} (14) \citet{1995ApJS...99..223P} 
(15) \citet{2016AJ....151...33G}
(16) \citet{2015MNRAS.448.2608V} (17) \citet{2017ApJ...848....5B} (18) \citet{2013MNRAS.436.3614S} (19) \citet{2014ApJ...781...69M} (20) \citet{2010ApJ...709.1343M} (21) \citet{2011MNRAS.413.2140T}  (22) \citet{1995A&AS..110..513B} (23) \citet{2000AJ....120.1487M} 
  (24) GALEXASC J202849.46-042255.5 (25) \citet{2020MNRAS.499..974G} (26) \citet{2016MNRAS.462..137T} (27) \citet{1981PASP...93...36D}} 

\renewcommand{\thetable}{A\arabic{table}}
\begin{tabular}{lcccccc} % four columns, alignment for each
		\hline
		SN Name & Type & Host & cz            & Distance$^1$  & E(B-V) & Reference(s)\\
                &&   Galaxy    &   (kms$^{-1}$)&  (Mpc) &  &   \\
		\hline
2023ixf & II-P & M101 & 241 $\pm$ 2 & 6.85 $\pm$0.15 & 0.039  & (2) \\
2020jfo & II-P & NGC 4303 & 1556 $\pm$ 2 & 	14.623 $\pm$ 1.957& 0.020  & (3) (4)\\
2018zd & II-P$^*$ & NGC 2146  & 892 $\pm$ 4 & 9.6 $\pm$ 1.0    &0.08   &    (5), (6)\\
2013ej & II-P & NGC 628 & 657 $\pm$ 1  & 9.7 $\pm$ 0.5  &0.06   & (7)\\ 
2012aw & II-P & M95 & $778\pm2$ & $9.9\pm0.1$ &  0.074 &   (8) \\
2004et & II-P & NGC 6946& 	48 $\pm$ 2   & 5.7 $\pm$ 0.3   &0.41 &  (9), (10), (11)\\

1999em & II-P & NGC 1637 &   670 $\pm$ 3 & 9.77 $\pm$ 1.82 &  0.04   & (12), (13)\\
1987A &  II-P & LMC & 278 $\pm$ 2 & 0.05 &  0.055& (14), (15) \\
2013by & II-P/L & ESO 138-G10 & 1144 $\pm$2 & $\sim$14.8 & 0.195 &  (16), (17) \\

2011dh & II-b & M51 &  	393   &8.05 $\pm$ 0.35   &    0.3 & (18), (19)\\
2008ax &II-b  & NGC 4490 & 565 $\pm$ 3 & 	5.979 $\pm$ 0.379  & 0.022 &  (20), (21)\\
1993J & II-b & NGC 3031  & -39 $\pm$ 3 & 3.675 $\pm$ 0.049 &  0.3 &    (22), (23)\\
2017ivv & II-L & (24)  & 1680 $\pm$ 200 & 24.09 $\pm$ 2.90 &  0.05 & (25)  \\
2014G & II-L & NGC 3448 & 1528 $\pm$ 13 & 22.53 $\pm$ 1.59 & 0.21 & (26) \\
1979C & II-L & M100 & $1571 \pm 1$ & $16.117 \pm0.365$ & 	0.072 & (27)\\
\end{tabular} \\  
$^*$Formally recognised as a Hydrogen rich electron-capture SN (5) which showed nebular spectroscopic features of a normal II-P SN
\label{Tab:SNeSampleData}
\end{table*}

Table \ref{Tab:SNeSampleData} outlines the sample of comparison CC-SNe used in this study.

% Don't change these lines
\bsp	% typesetting comment
\label{lastpage}
\end{document}